\documentclass[journal]{IEEEtran}
\ifCLASSINFOpdf
\else
\fi
\usepackage{url}
\usepackage{amsmath,amssymb,amsfonts}
\usepackage{algorithmic}
\usepackage{graphicx}
\usepackage{textcomp}
\usepackage{multirow}
\usepackage{mathtools}
\usepackage{booktabs}
\usepackage{xcolor}
\usepackage[caption=false]{subfig}
\usepackage{xifthen}
\usepackage{nicefrac}
\usepackage{fancyhdr}

\usepackage[draft]{todonotes}

\definecolor{my_red}{rgb}{0.7,0,0}
\definecolor{my_blue}{rgb}{0,0,0.7}

\newtheorem{definition}{Definition}

\newcommand{\x}{\mathbf{x}}
\newcommand{\C}{\mathbf{C}}
\newcommand{\RN}{\mathcal{I}}
\newcommand{\F}{\mathcal{F}}
\newcommand{\Fell}{F_{-\ell}}

\newcommand{\DCT}{\textbf{DCT}}
\newcommand{\META}{\textbf{META}}
\newcommand{\HEADER}{\textbf{HEADER}}

\newcommand{\lab}[3]{
    \ifthenelse{\isempty{#3}}{\ifthenelse{\isempty{#2}}{\textrm{#1}}{(\textrm{#1},\textrm{#2})}}{(\textrm{#1},\textrm{#2},\textrm{#3})}
}

\fancypagestyle{firstpage}
{
    \fancyhead[C]{\textit{This work has been submitted to the IEEE for possible publication. Copyright may be transferred without notice, after which this version may no longer be accessible.}}    
}

\begin{document}
\title{Multi-clue reconstruction of sharing chains \\ for social media images}

% \author{Author1,~\IEEEmembership{Member,~IEEE,}
%         Author2,~\IEEEmembership{Fellow,~OSA,}
%         and~Author~3,~\IEEEmembership{Life~Fellow,~IEEE}% <-this % stops a space
% \thanks{M. Shell was with the Department
% of Electrical and Computer Engineering, Georgia Institute of Technology, Atlanta,
% GA, 30332 USA e-mail: (see http://www.michaelshell.org/contact.html).}% <-this % stops a space
% \thanks{J. Doe and J. Doe are with Anonymous University.}% <-this % stops a space
% \thanks{Manuscript received April 19, 2005; revised August 26, 2015.}}
\author{Sebastiano Verde,
        Cecilia Pasquini,
        Federica Lago,
        Alessandro Goller,\\
        Francesco De Natale,
        Alessandro Piva,
        and~Giulia Boato%
\thanks{S. Verde, C. Pasquini, F. Lago, A. Goller, F. De Natale and G. Boato are with the \textit{Dipartimento di Ingegneria e Scienza dell'Informazione} (DISI), University of Trento, Italy.}%
\thanks{A. Piva is with the \textit{Dipartimento di Ingegneria dell'Informazione} (DINFO), University of Florence.}%
\thanks{Manuscript received August 4, 2021; revised Month DD, YYYY.}}       

\maketitle

\thispagestyle{firstpage}
\begin{abstract}
The amount of multimedia content shared everyday, combined with the level of realism reached by recent fake-generating technologies, threatens to impair the trustworthiness of online information sources. The process of uploading and sharing data tends to hinder standard media forensic analyses, since multiple re-sharing steps progressively hide the traces of past manipulations. At the same time though, new traces are introduced by the platforms themselves, enabling the reconstruction of the sharing history of digital objects, with possible applications in information flow monitoring and source identification. In this work, we propose a supervised framework for the reconstruction of image sharing chains on social media platforms. The system is structured as a cascade of backtracking blocks, each of them tracing back one step of the sharing chain at a time. Blocks are designed as ensembles of classifiers trained to analyse the input image independently from one another by leveraging different feature representations that describe both content and container of the media object. Individual decisions are then properly combined by a late fusion strategy. Results highlight the advantages of employing multiple clues, which allow accurately tracing back up to three steps along the sharing chain.

\end{abstract}

% \begin{IEEEkeywords}
% IEEE, IEEEtran, journal, \LaTeX, paper, template.
% \end{IEEEkeywords}

% \fi
\IEEEpeerreviewmaketitle

% *** SECTIONS ***

\section{Introduction}\label{sec:introduction}

\IEEEPARstart{M}{assive} amounts of multimedia data are uploaded every day to social media platforms by nearly 4 billion active users: according to recent estimates, 3.2 billion images are shared every day~\cite{smith2019} and 500 hours of video are uploaded to YouTube every minute~\cite{statista2021}. At the same time, easy-to-use editing tools that allow modifying such multimedia data have become widely available. While this enables for an unprecedented ease in sharing information, it also entails serious implications for the trustworthiness and reliability of digital media. 

Such concerns reached a critical level with the recent development of tools based on artificial intelligence (AI) that allow even inexperienced users generating almost automatically highly realistic fakes, especially when dealing with images and videos depicting faces~\cite{thies2016face,ngo2021self}. Indeed, advanced tools like AI and photo/video editing used to be restricted to skilled users and researchers but are nowadays available to a much wider public, likely going beyond the primary purpose of entertainment. As for any other technology, a reasonable risk exists for malicious misuse, e.g., conveying misinformation to bias people and influence social groups~\cite{allen2017}.
Moreover, multimedia data are responsible for the viral diffusion of information through social media and web channels, and play a key role in the digital life of individuals and societies. Therefore, developing tools to preserve the trustworthiness of shared images and videos is a necessary step that our society can no longer ignore.

Several works in media forensics have investigated the detection of manipulations and the identification of the digital source, providing promising results in laboratory condition and well-defined scenarios \cite{Verdoliva2020media}. More recently, the research community has also pursued the ambition to scale forensic analyses to real-world web-based systems, which involve routinely applied operations such as the act of sharing through social media platforms~\cite{pasquini2021media}. This extension requires the ability to face significant technological challenges related to the (possibly multiple) uploading/sharing processes, and hence the need for methods that can reliably work under more general conditions. Retrieving information about the life of a digital object in terms of provenance, manipulations and sharing operations, would indeed represent a valuable asset: on one hand, it could support law enforcement agencies and intelligence services in tracing perpetrators of deceptive visual contents; on the other hand, it could help in preserving the trustworthiness of digital media and countering the effects of misinformation.

The process of uploading to web platforms represents nowadays a key phase in the life of a digital object, and the dynamics of shared visual content can be analyzed for different purposes~\cite{cheung2015connection,mada2019trust}. While this sharing process typically hinders the ability to perform conventional media forensics tasks, it also introduces new traces itself, allowing to infer additional information. As a matter of fact, data can be uploaded in different ways, multiple times, on diverse platforms and from different systems. In this context, the possibility of reconstructing the sharing history of a given object, known as \emph{platform provenance analysis}~\cite{pasquini2021media}, could help monitoring the information flow by tracing back previous uploads, and thus supporting source identification by narrowing down the search.

% In recent years such possibilities were the focus of several works, that we gather under the name of \emph{platform provenance analysis} \cite{pasquini2021media}. We can broadly summarize their goals as follows:
% \begin{itemize}
%     \item 	Identification of the platforms that processed the object;
%     \item	Reconstruction of the full sharing history of the object;
%     \item 	Detection of the systems used in the upload phase.
% \end{itemize}\todo[inline]{Questo paragrafo non mi è chiaro messo li cosi... ci serve citare le tre direzioni oppure no? Per me o si toglie (citando la survey sopra) oppure bisogna legarlo meglio al resto e capire cosa vogliamo dire.}

\begin{figure*}
    \centering
    \includegraphics[width=\textwidth]{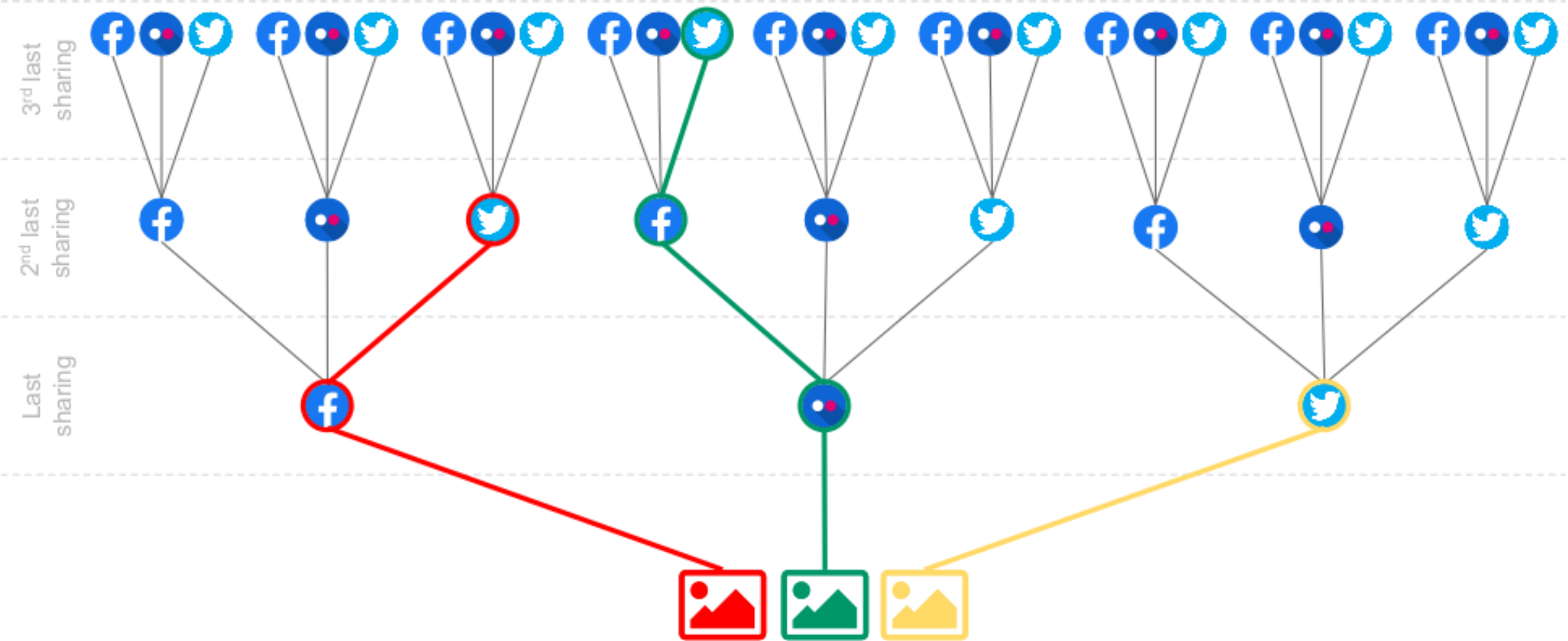}
    \caption{Visual representation of the media recycling problem where we aim at reconstructing the sharing history of a given image. In this work we consider up to three sharing steps on three different platforms, namely Facebook (FB), Flickr (FL) and Twitter (TW). Three examples of reconstruction of the sharing history are presented: the red one has been shared first on TW and then on FB; the green one on TW, FB and FL; the yellow only on TW (note that the reconstruction proceeds backwards, starting from the most recent sharing step).}
    \label{fig:sharing}
\end{figure*}

Distinct traces are left on a digital image when uploaded to a web platform or a social network, depending on the operations that are involved in the process. As firstly observed in the case of Facebook~\cite{moltisanti2015image}, compression and resizing are typically applied to reduce the size of uploaded images, and this is performed differently on different platforms and depending on the resolution of the original image. As known in media forensics, such operations can be detected and characterized by analyzing the image content or signal (i.e., the values in the pixel domain or in various transformed domains). A signal-based approach for platform provenance analysis can leverage the Discrete Cosine Transform (DCT) coefficients~\cite{caldelli2017, amerini2017}, the PRNU noise~\cite{caldelli2018} or combinations of the two~\cite{amerini2019social}. Moreover, provenance can also be inferred as a by-product of detectors developed to analyse image manipulations in the pixel domain, as shown in~\cite{mazumdar2019}.

Meaningful information can be extracted as well from the image container, such as metadata. While signal-based forensic analysis is usually preferable (as data structures can be erased or falsified more easily than signals), such clues play a relevant role in platform provenance analysis, being typically related to the platform itself rather than to the acquisition device~\cite{mullan2019}. In~\cite{giudice2017} the authors consider several popular platforms (namely Facebook, Google+, Flickr, Tumblr, Imgur, Twitter, WhatsApp, Tinypic, Instagram, Telegram), observing two main facts: (i) the uploaded files are renamed with distinctive patterns which can even allow to reconstruct the file's web URL; and (ii) both resizing and JPEG compression adopt platform-specific rules. Therefore, the authors propose a feature representation that includes image resolution and quantization table coefficients, which can be extracted from the file without decoding.

An even more challenging scenario consists in the reconstruction of sharing chains, meaning that the target of the analysis goes beyond the identification of the latest sharing platform and aims at reconstructing the history of a digital object that has been re-shared multiple times, possibly on different platforms. We denote this scenario as \emph{media recycling}. Since forensic traces tend to decay as we keep applying new operations on an image, the problem of media recycling requires the combination of multiple detection strategies to be solved. Hybrid approaches where clues are extracted from both signal and metadata have been proposed in~\cite{phan2018,phan2019,siddiqui2019}, showing promising results in the identification of sharing steps beyond the most recent one.

In this paper we address the image recycling problem by proposing a novel multi-clue detection system able to reconstruct the sharing history of images on various platforms. The proposed framework is designed as a cascaded architecture, which allows tracing back one sharing platform at a time, leveraging the knowledge of the previously detected steps to reconstruct the whole sequence step by step (Figure~\ref{fig:sharing}). The system employs an effective fusion mechanism to combine multiple classifiers, thus allowing to successfully exploit both content and container of the inspected image, and being open to possible integration with other sets of features. As an additional contribution, we present a novel set of container-related features extracted from the JPEG header, which allows significantly boosting performances when combined to other detectors. 

%This aspect is also explored in \cite{mullan2019}, where the authors aim at linking JPEG headers of images acquired with Apple smartphones and shared on different apps to their acquisition device; their analysis show that JPEG headers can be used to identify the operating system version and the sharing app used to a certain extent.

%To combine the analysis results from different recycling traces, we implemented a fusion mechanism based on the BKS paradigm~\cite{Huang1993} that models the problem as a combination of multiple ``experts'', estimating a probability map from the set of outputs of the individual classifiers to the overall decision. The step-by-step reconstruction of sharing chains is then obtained through a cascaded architecture, where the classification at a given step is driven by the detection results at the previous ones. \todo[inline]{Anche questo secondo me è troppo dettagliato per stare in intro... possiamo anche qui volendo snellire e completare il paragrafo sopra ma non mi metterei a spiegare il BKS qui... per altro va eplicitato acronimo la prima volta che lo nominiamo.}

The proposed framework is evaluated on a published dataset of images shared on different platforms. Experiments, conducted including both single detectors and fused ones, show that the combination of heterogeneous traces is beneficial in media recycling detection. Thanks to the novel set of container-based features, the identification of the last sharing step reaches 100\% accuracy on the test data. Moving further back in the history, the system allows reconstructing the sharing chain with high accuracy up to the third step.  

A repository with the implementation of the presented system is publicly available\footnote{\url{https://github.com/Flake22/sharing-chains-reconstruction}}.

The paper is structured as follows: Section~\ref{sec:multi} formally defines the image recycling problem and describes the architecture of the proposed reconstruction system; Section~\ref{sec:feat} presents a set of forensic traces that provide meaningful information in the context of the image recycling scenario, along with the adopted strategy for fusing such multiple descriptors; Section~\ref{sec:results} discusses the experimental setting and the obtained results, and reports a feature separability analysis for the presented descriptors; finally, Section~\ref{sec:conc} draws the conclusions.
\section{Multi-step reconstruction of sharing chains}
\label{sec:multi}

The problem of image recycling involves media data that have been shared one or more times through (possibly different) social media and web platforms. 

The key assumption is that a given image $\x \in \RN$ underwent a number $\ell\geq 1$ of sharing steps, forming a {\it sharing chain}. 

\begin{definition}
Given a set $\mathcal{S}$ of sharing platforms, a \emph{sharing chain} of length $\ell$ is a sequence of sharing platforms, which we can be represented as a vector $\C$ in $\mathcal{S}^\ell \doteq \underbrace{\mathcal{S} \times \ldots \times \mathcal{S}}_\ell$. 
\end{definition}

For the sake of convenience, we will index the components of $\C$ in reverse order, so that the temporal succession of sharing platforms goes as follows: 
\begin{equation}\label{eq:chain}
    \C[-(\ell-1)] \to \ldots \to \C[-1] \to \C[0],
\end{equation}
and thus $\C[0]$ corresponds to the last sharing step in $\C$.

\begin{definition}
For a generic chain $\C \in \mathcal{S}^\ell$, the set $\mathcal{B}(\C)$ contains all the chains of length $\ell+1$ whose last $\ell$ components are identical to the ones of $\C$; in other words, chains in $\mathcal{B}(\C)$ differ from $\C$ by one additional previous sharing step, $\C[-\ell]$.
\end{definition}

Let us also define $\Omega_\ell$ as the set of all possible sharing chains of length up to $\ell$ obtained through the combination of platforms in $\mathcal{S}$, 
\begin{equation}
    \Omega_\ell \doteq \bigcup_{1\leq i\leq\ell} \mathcal{S}^i
\end{equation}

In our formulation, the goal of a recycling analysis is to devise a system $\F_L$ that assigns a certain image $\x$ to the sharing chain it went through, up to a predefined maximum chain length~$L$.   

% As described in Section \ref{sec:multi}, in our work $\F_L$ is a supervised system that classifies an input image by exploiting a multi-domain feature representation. In particular, for each input sample $\x$ three different feature vectors are extracted, namely {\DCT}, {\META}, and {\HEADER}, which are described in detail in Section \ref{sec:feat}.

\subsection{Cascaded architecture}\label{ssec:cascade}

Given a predefined number $L$ of sharing steps, our purpose is to devise a system $\F_L: \RN \rightarrow \Omega_L$ that takes in input an image $\x$ and associates it to the correct chain of maximum length $L$.

In our approach, we propose to structure $\F_L$ as a multi-step cascade of {\it backtracking blocks} $\Fell$, each of them tracing back one step of the sharing chain at a time. 

In particular, we define:
\begin{itemize}
    
    \item $F_0 : \RN \rightarrow \Omega_1 \equiv \mathcal{S}$
    
    This first backtracking block assigns $\x$ to the platform in $\mathcal{S}$ that corresponds to the last sharing step.
    
    \smallskip

    \item $\Fell : \RN \times \Omega_\ell \rightarrow \Omega_{\ell+1}, \quad \ell=1, \ldots, L-1$

    For $\ell>1$, based on the knowledge of the previous blocks' decisions, each backtracking block inspects a possible preceding sharing operation. This process is recursively performed until $\ell$ reaches $L-1$, and the last block provides the final output. 
    
\end{itemize}

Formally, we can express the whole system as
\begin{equation}\label{eq:comp}
    \F_L(\x) = F_{-(L-1)}(\x, \ldots , F_{-2}(\x,F_{-1}(\x, F_0(\x))\ldots )
\end{equation}

\begin{figure}[ht]
    \centering
    \includegraphics[width=\columnwidth]{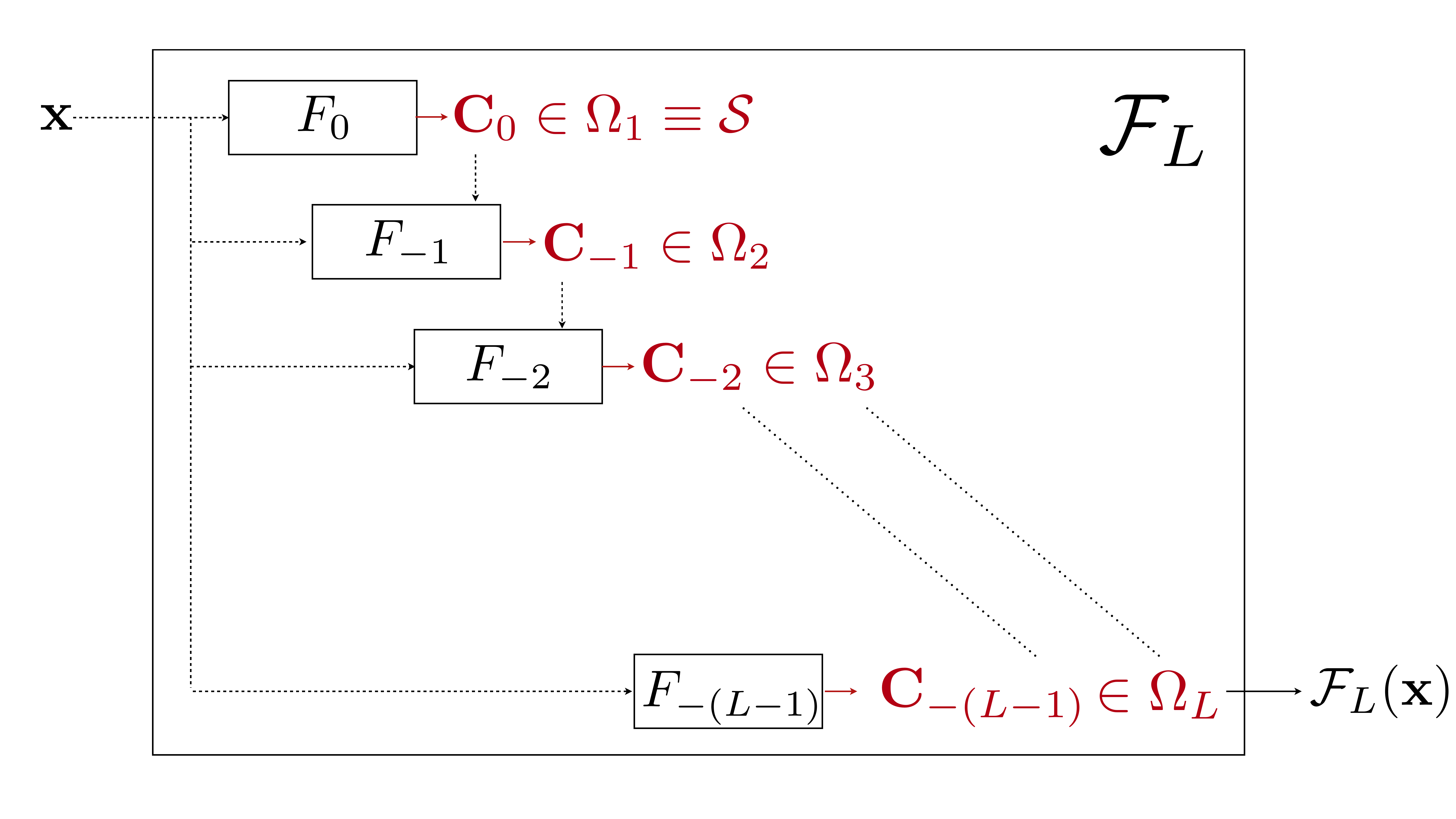}
    \caption{Schematic representation of the multi-step cascade. Backtracking blocks are represented in squares; black dashed lines indicate input arguments to the blocks, solid red lines the output of the blocks.}
    \label{fig:scheme}
\end{figure}

By defining the intermediate chains $\C_{-\ell}$ as the output of $F_{-\ell}$, i.e., $\C_{-\ell} \doteq \Fell(\x,F_{-(\ell -1)}(\x))$, we can represent the multi-step process as in Figure \ref{fig:scheme}.

Each intermediate backtracking block assigns $\x$ to a potentially longer chain, in the case a previous sharing step is detected. This is done through an ensemble of classifiers, each one specialized in dealing with a specific output of the previous block.
If no additional steps are detected, the intermediate chain does not increase in length and is regarded as the final output. 
By dropping the subscript for the sake of simplicity and indicating as $\C \in \Omega_\ell$ an arbitrary intermediate chain determined at the previous backtracking block, we can formulate a generic intermediate block $\Fell$ as follows:

\begin{equation}\label{eq:block}
    \Fell(\x,\C) = 
    \begin{cases}
    \C & \text{if } \C \in \Omega_{\ell-1} \\
    f^\C_{-\ell}(\x)& \text{if } \C \in \Omega_\ell \setminus \Omega_{\ell-1}
    \end{cases}
\end{equation}

The first case corresponds to the situation where no backward step is detected at the previous backtracking block (i.e., the length of $\C$ is lower than $\ell$), therefore no additional steps should be added to the current chain. In the second case, 
the functions $f^\C_{-\ell}(\cdot)$ are specialized detectors trained to disambiguate among the set containing the chain $\C$ and all the ones that include a previous sharing step. Thus, by design, one detector
$f^\C_{-\ell}:\RN \rightarrow \{\C\} \cup \mathcal{B}(\C)$ is needed for each $\C \in \Omega_\ell \setminus \Omega_{\ell-1}$. By design, we indicate $f_0(\x) \equiv F_0(\x)$.

The way specialized detectors assign $\x$ to either $\C$ or a longer chain involves the combination of different recycling traces. The description of the employed traces and the fusion technique adopted to combine the associated detectors is the subject of the next section.

\begin{figure}
    \centering
    \includegraphics[width=\columnwidth]{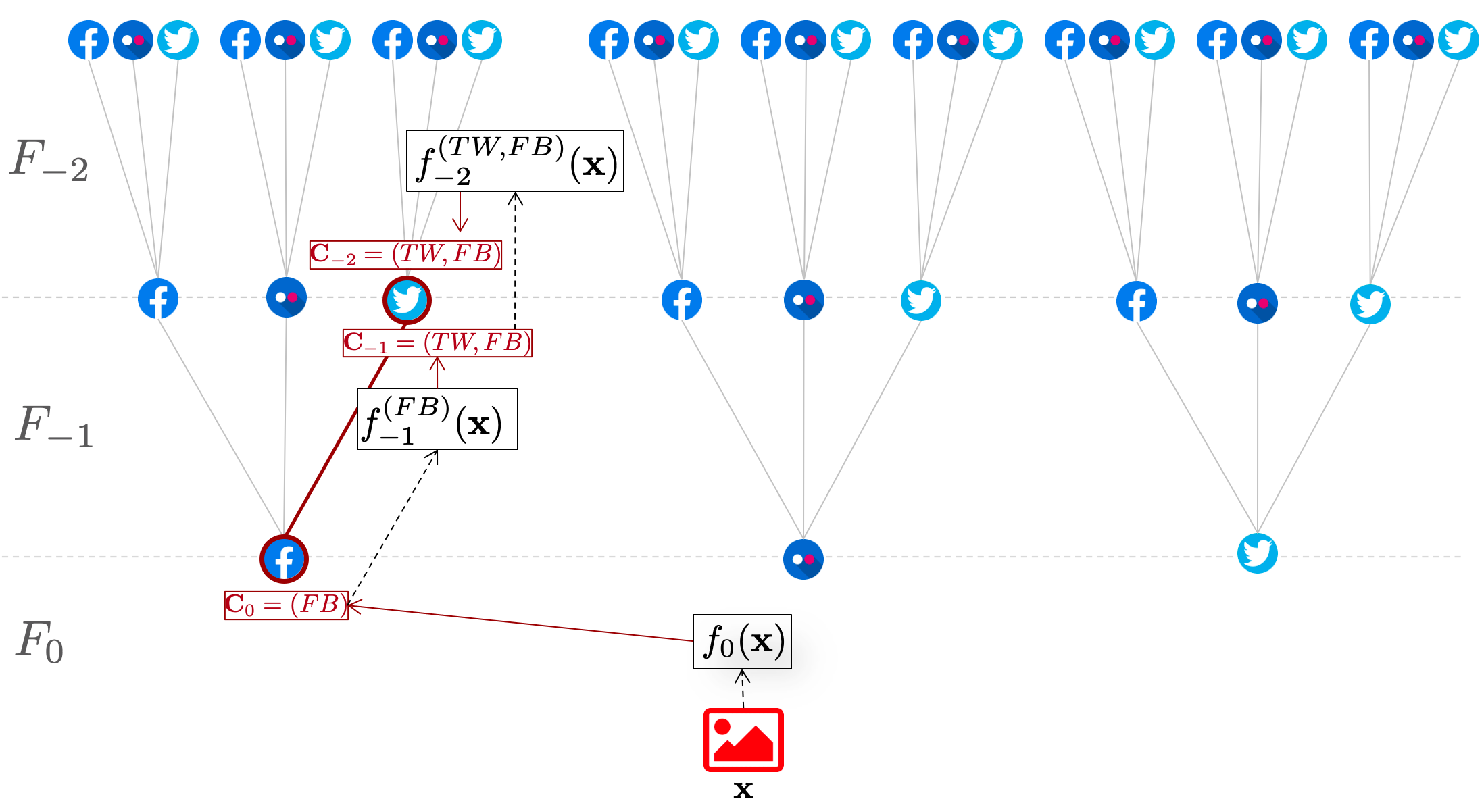}
    \caption{Visual example of a run of the system $\mathcal{F}_L$, with $L=3$ and $\mathcal{S}=\{\text{FB}, \text{TW}, \text{FL}\}$. Backtracking blocks are annotated in gray on the left; black dashed lines represent arguments that are given in input to specialized detectors, while solid red lines indicate the outputs of specialized detectors.
    In this case, the chain $(\text{TW},\text{FB})$ is reconstructed: after $F_0(\x)$ identifies Facebook as last sharing platform $\C_0$, the block $F_{-1}(\x,\C_0)$ calls the specialized detector $f_{-1}^{(\text{FB})}(\x)$, which detects a previous sharing on Twitter and returns the chain $\C_{-1}=(\text{TW},\text{FB})$; then, the block $F_{-2}(\x,\C_{-1})$ calls the specialized detector $f_{-2}^{(\text{TW},\text{FB})}(\x)$, which returns again the chain $\C_{-2}=(\text{TW},\text{FB})$, thus fixing the chain length to $2$. The final output of the system is therefore $\mathcal{F}_{3}(\x)=\C_{-2}=(\text{TW},\text{FB})$.}
    \label{fig:esempio}
\end{figure}

For the sake of clarity, we report in Figure \ref{fig:esempio} a visual example of how the system $\mathcal{F}_L$ works for the case $L=3$ and $\mathcal{S}=\{\text{FB}, \text{TW}, \text{FL}\}$, which refer to Facebook, Twitter and Flickr, respectively.

\section{Extraction and fusion of recycling traces}
\label{sec:feat}

Traces of media recycling can be found on image data under investigation exploiting different domains. In our case, we define multiple feature representations extracted from both image signal and image container.
As far as signal-based features are concerned, we focus on the histograms of DCT coefficients (denoted as {\DCT} and described in Section~\ref{ssec:content}), which have been successfully adopted in the literature to address several media forensics problems~\cite{phan2018,ih2016,pevny2008detection}. 
Regarding container-based information, two different feature vectors are defined: {\META}, which encodes metadata mostly related to the last JPEG compression settings, and {\HEADER}, which contains information on the overall structure of the JPEG header; these feature descriptors are described in Section~\ref{ssec:container}.
In particular, the latter represents a recycling trace that we propose for the first time in this work and is capable of boosting the overall performance when combined to other features.

% *** Content-based ***

\subsection{Content-based features}\label{ssec:content}

The set of content-based features, denoted as {\DCT}, encodes the forensic traces left on the image signal by the lossy compression of JPEG standard, which is currently the most common format for images uploaded on social networks. The specific processing, however, is peculiar to the needs of each platform. In particular, the target image quality is controlled by the integer quantization in the DCT domain, which is reflected in the statistics of the quantized coefficients. 

Following the scheme in~\cite{caldelli2017}, the $8\times 8$ block-based DCT is first computed on the whole image; then, normalized histograms of dequantized DCT coefficients are extracted from the first 9 AC frequencies, in zigzag order. We retained for each histogram the $41$ bins corresponding to the range of integers between $-20$ and $20$. Finally, the concatenation of the histograms provides a 369-dimensional feature vector.

% Figure~\ref{fig:dct} shows some examples of histograms of DCT coefficients for the same image in different sharing scenarios. One can observe how differences in DCT statistics are present both among different platforms (Figure~\ref{fig:dct}a) and when we compare a single upload (in the example on Facebook) to a recycling of the image from a different platform (Figure~\ref{fig:dct}b).

% \begin{figure}
%     \centering
%     \subfloat[{Single share, $\C[0]$.}]{\includegraphics[width=\columnwidth]{figures/C0.jpg}}\\
%     \subfloat[{Re-sharing, $(\C[-1],\textrm{FB})$}.]{\includegraphics[width=\columnwidth]{figures/C1.jpg}}
%     \caption{Examples of histograms of DCT coefficients for the same image shared once on each platform (a) and recycled (b).}
%     \label{fig:dct}
% \end{figure}

% *** Container-based ***

\subsection{Container-based features}\label{ssec:container}

% *** META ***

The first set of container-based features, denoted as {\META}, exploits the information about the JPEG compression settings contained in the metadata of the image under investigation. These features, as proposed in \cite{phan2019}, consist of a 152-dimensional vector encoding the following information:
\begin{itemize}
    \item \textit{Quantization tables} (128), for luminance and chrominance channels, reflecting the JPEG quality factor
    \item \textit{Huffman encoding tables} (2), number of tables used for AC and DC component
    \item \textit{Component information} (18), describing component id, horizontal/vertical sampling factors, quantization table index and AC/DC coding table indices, for each YCbCr component
    \item \textit{Optimized coding and progressive mode} (2), binary features indicating the use of the two modes
    \item \textit{Image size} (2), the image dimensions
\end{itemize}

% *** HEADER ***

The second set of features, denoted as {\HEADER}, is a novel contribution of this work and encodes structural properties of the JPEG header.%, which is divided into segments each of them starting with a marker.

Previous approaches in~\cite{farid2009,mullan2019} observed that the EXIF information of JPEG files can contain useful information for forensics purposes, such as authentication and source identification. More recently, authors in~\cite{yang2020efficient} proposed an efficient container-based method to verify the integrity of videos, showing also the possibility to identify the social media platform on which the video was uploaded. % The advantages of this method are numerous, being it computationally efficient due to the low feature dimensionality and independent of the size of the video, as information are only extracted from the file header.  

A similar idea is exploited in this work, adapted to the image recycling scenario. In fact, while sharing platforms usually strip out optional metadata fields (like acquisition time and device, GPS coordinates, etc.), we observed that different platforms retain different EXIF fields in the downloaded JPEG files, which can be extracted through several possible tools and be employed as feature descriptors. 

\begin{figure}[t]
    \centering
    \includegraphics[width=\columnwidth]{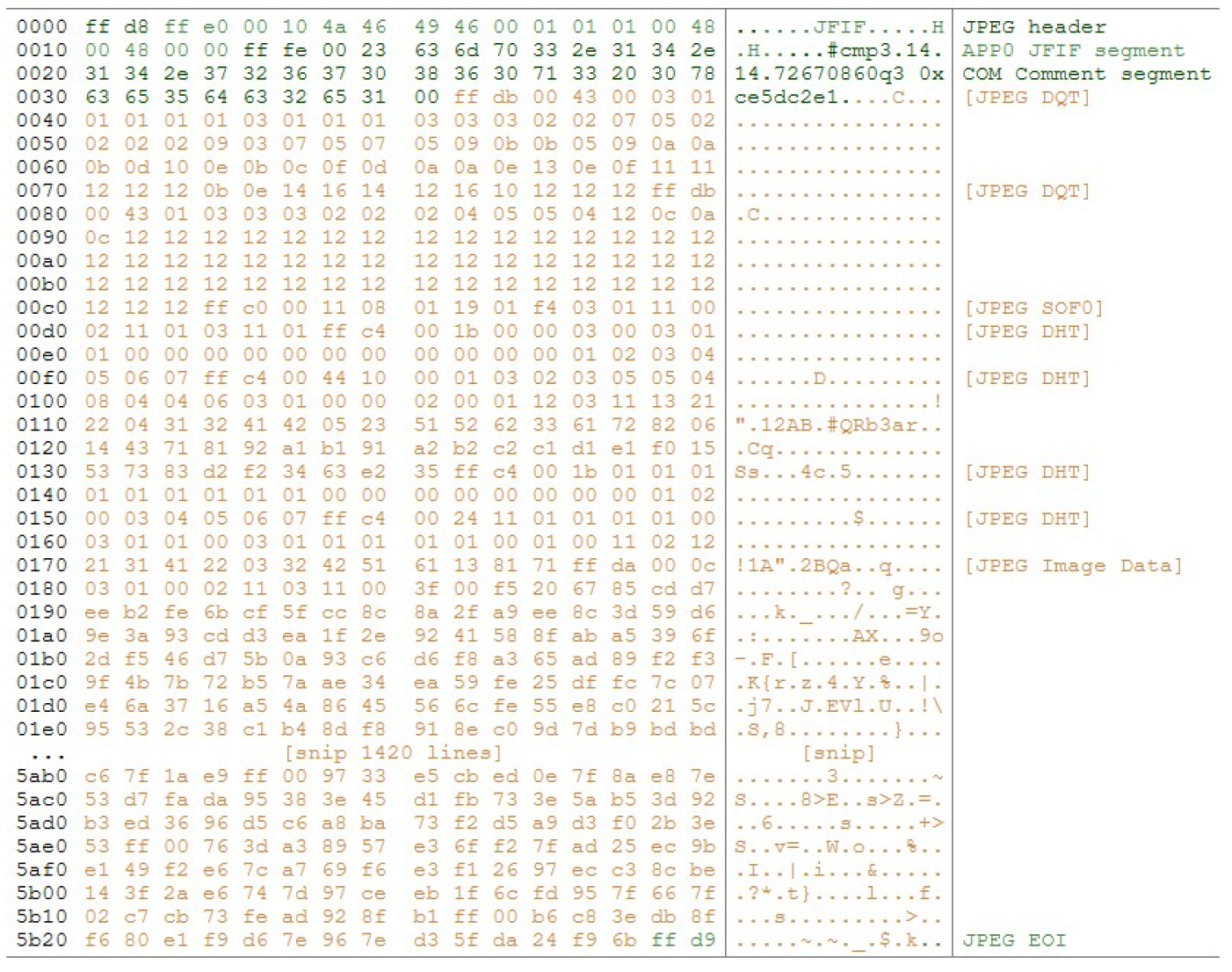}
    \caption{Example of JPEG header representation as extracted by ExifTool. In this case, the image has been shared once through Twitter.}
    \label{fig:header}
\end{figure}

Our implementation is based on ExifTool~\cite{exiftool}, a free and open-source software library for reading, writing, and editing image containers.
%Exiftool, respect to other software, implements its own open metadata format. It is designed to encapsulate meta information from many sources, in binary or textual form, and bundle it together with any type of file [4]. 
In particular, ExifTool encodes the header of a JPEG file into an HTML page, as exemplified in Figure~\ref{fig:header}. 

%The first column is the number of hexadecimal elements, the second column is the hexadecimal code, the third column is the ASCII version of the hexadecimal code, and the last column is the JPEG signature format, the header of the image.

Every extracted file contains header markers indicating the beginning of a specific kind of segment. Those can be found multiple times throughout the header, and their frequency represents a discriminative property for detecting the upload on a sharing platform.

For this purpose, a set of 8 markers was selected, as detailed below:
\begin{itemize}
    \item \textit{DHT}: encapsulates the information regarding the Huffman Table
    \item \textit{unused}: unused data blocks for maintaining a fixed size
    \item \textit{APP13}: provides a number of methods for managing Photoshop/IPTC data without dealing with the details of the low level representation
    \item \textit{APP2}: used to store the ICC profile
    \item \textit{SOF0}: Start Of Frame (Baseline DCT), indicates a baseline DCT-based JPEG, and specifies the width, height, number of components, and component subsampling
    \item \textit{SOF2}: Start Of Frame (Progressive DCT), indicates a progressive DCT-based JPEG, and specifies the width, height, number of components, and component subsampling
    \item \textit{cmp3}: comment section
    \item \textit{JPEG DRI}: Define Restart Interval.
\end{itemize}

\begin{figure}[t]
    \centering
    \includegraphics[width=\columnwidth]{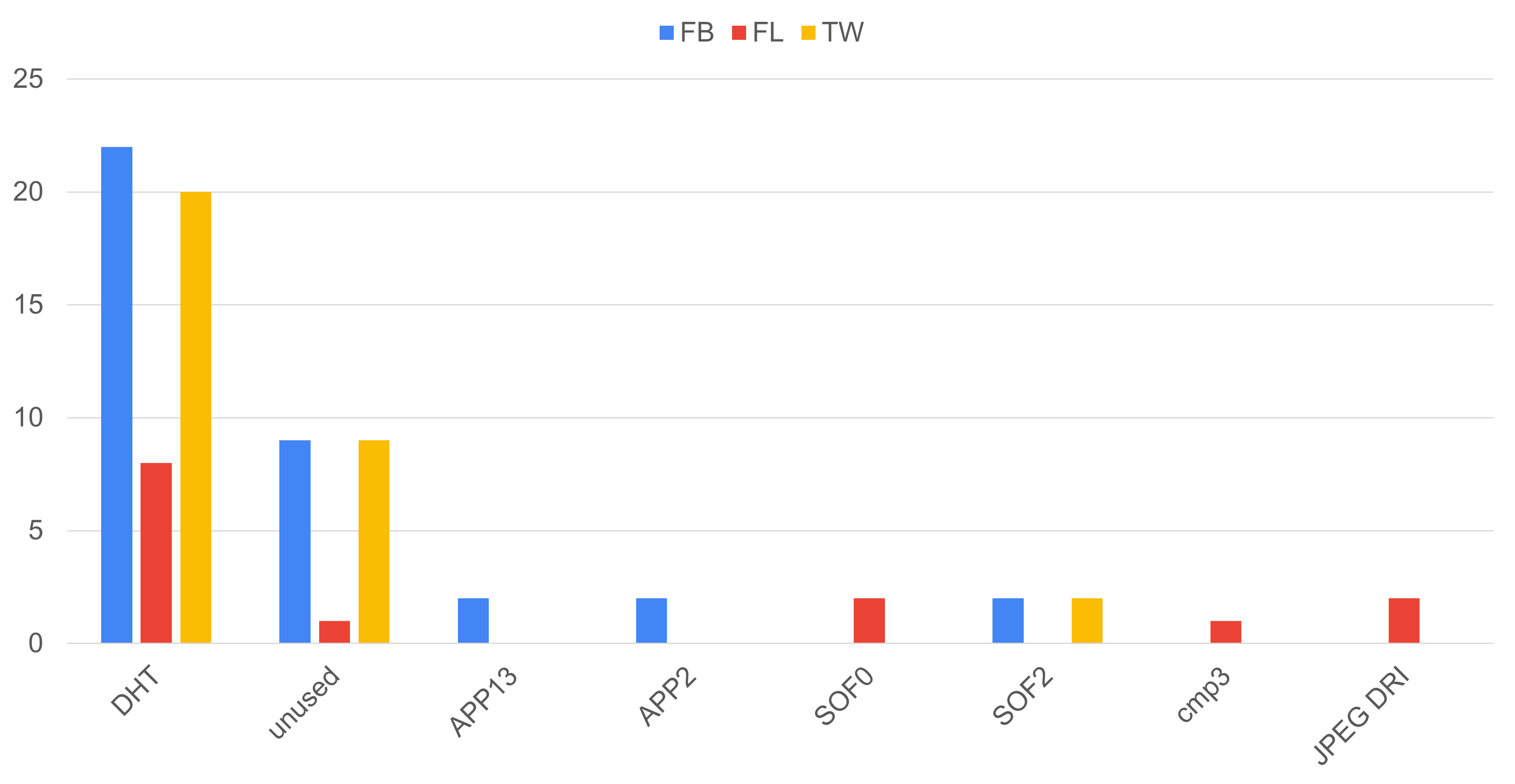}
    \caption{Comparison of {\HEADER} features extracted from the same image shared on different platforms.}
    \label{fig:headerFeatures}
\end{figure}

For each image file, the frequency of the listed segments is computed throughout the JPEG header, providing a 8-element feature vector. An example is reported in Figure~\ref{fig:headerFeatures}, where the feature vector is extracted for the same image when subject to a sharing operation on different platforms. The resulting {\HEADER} descriptor is therefore low-dimensional and does not require to decode the image, since information are read directly from the file header. In addition to being computationally efficient, {\HEADER} provides an extremely accurate identification of the last sharing step, and remains informative enough to boost the performance of other descriptors in further steps of the chain, as demonstrated in Section~\ref{sec:results}.    

%not decode the whole image, but instead only read out the header information, they are extremely light-weight, and can be used to rapidly sort coarsely huge amounts of les by camera model or make [13]. Only few works on images use this method to determine image provenance (CITARE), this is due to the ease of manipulate the header information, but until now, no one use it for tracing the social network posting history of an image.

% *** BKS FUSION ***

\subsection{Fusion of classifiers}\label{ssec:fusion}

\def\j{y}
\def\nunit{Q}
\def\iunit{q}

The described feature representations can be employed to train specific classifiers that discriminate among pre-defined sets of sharing chains. However, in order to implement the functions $f^\C_{-\ell}(\x)$ described in \eqref{eq:block}, we need to combine the output of such classifiers into a single overall decision. For easier reading, from now on we will simplify the notation by removing any reference to the block index $\ell$ and the previous output $\C$, so that we can discuss the implementation of a generic function $f(\x)$. 

The above scenario is typically referred to in the literature as \textit{combination of multiple experts} (CME)~\cite{Ruta2000,Moreno2006}, where each classifier is regarded as an ``expert'' who analyses one specific aspect of the object under investigation, and the final response is obtained by properly merging all the individual decisions.

Formally, the CME problem involves $K$ experts (or classifiers) denoted by $e_k$, $k=1,\ldots,K$, which share the same set of mutually exclusive output classes. 
% Let us thus define:
% \begin{itemize}
%     \item $e_D(\x)$, a DCT-classifier
%     \item $e_M(\x)$, a META-classifier
%     \item $e_H(\x)$, a HEADER-classifier
% \end{itemize}
The function $f(\x)$ combines the individual decisions (the outputs of the $K$ classifiers) by means of a fusion function $g$, and provides an overall classification,
\begin{equation}
    f(\x) = g\left(e_1(\x),e_2(\x),\ldots,e_K(\x)\right)
\end{equation}

Note that the individual decisions and the combined one all belong to the same set of chains (which depends on the block index and the result of the previous block). Regardless of the specific set, however, the number of chains is always $\vert\mathcal{S}\vert + 1$, since we can either add a new step from $\mathcal{S}$ to the partial chain or not.  

The CME problem, i.e., the implementation of the fusion function $g$, is addressed in our work by means of a method known as \textit{behavior-knowledge space} (BKS) \cite{Huang1993,Huang1995}, which allows building prior knowledge about the behavior of the individual classifiers without requiring the statistical independence of the same.
%As discussed in \cite{Huang1993}, if there is no prior knowledge about the behavior of the individual classifiers, then it is safer to treat them on an equal basis; however, when such information exists, better combination methods can be explored. Most CME methods though require an independence assumption on the classifiers, which usually does not hold in real applications~\cite{Huang1995}. To prevent this problem, in this work we implemented a solution known as \textit{behavior-knowledge space} (BKS) \cite{Huang1993,Huang1995}, which allows to build prior knowledge about the system without requiring the independence of classifiers.

A BKS is a $K$-dimensional space where each dimension is related to the decisions of one classifier. Each classifier has the same number of decisions, chosen from a given set, and the intersection of the decisions of individual experts identifies a \textit{unit} of the BKS. 

Given an input $\x$, for which we have $e_k(\x)$, $k=1,\ldots,K$, i.e., the individual decisions, the corresponding unit in the BKS has coordinates $(e_1(\x),e_2(\x),\ldots,e_K(\x))$ and is called the \textit{focal unit} for $\x$. Given that, in our case, the number of output classes for each classifier is $\vert\mathcal{S}\vert + 1$, it follows that the BKS contains exactly $\nunit=(\vert\mathcal{S}\vert + 1)^K$ units, denoted by $u_\iunit$, $\iunit=1,2,\ldots,\nunit$.

In practice, the BKS is implemented as a lookup table that associates each unit (i.e., each combination of individual decisions) to a probability distribution over the set of output classes. Such distributions are estimated by accumulating the number of incoming samples for each class on a dedicated training set (different from the one used to train the individual classifiers~\cite{Raudys2003}).

Let us define $y\in\{1,2,\ldots,\vert\mathcal{S}\vert + 1\}$, a set of numerical labels associated to the output classes. A representation of a BKS lookup table is given in Table~\ref{tab:bks_table}: each unit $u_\iunit$, $\iunit=1,\ldots,Q$, corresponds to a unique combination of classifier decisions, and $n_y^{(q)}$ is the number of training samples belonging to class $y$ that have fallen into the $q$-th unit.

{\setlength{\tabcolsep}{7pt}
\begin{table}[ht]
    \centering
    \caption{BKS lookup table.}
    \label{tab:bks_table}
    \begin{tabular}{ccccccc}
        \toprule
        ~ & \multicolumn{6}{c}{BKS units} \\
        ~ & $u_1$ & $u_2$ & $\ldots$ & $u_\iunit$ & $\ldots$ & $u_\nunit$ \\
        \midrule %\cline{2-6}
        \parbox[t]{3mm}{\multirow{4}{*}{\rotatebox[origin=c]{90}{Class $y$}}} & $n_1^{(1)}$ & $n_1^{(2)}$ & $\ldots$ & $n_1^{(\iunit)}$ & $\ldots$ & $n_1^{(\nunit)}$ \\
        ~ & $n_2^{(1)}$ & $n_2^{(2)}$ & $\ldots$ & $n_2^{(\iunit)}$ & $\ldots$ & $n_2^{(\nunit)}$ \\
        ~ & $\vdots$ & $\vdots$ &  & $\vdots$ &  & $\vdots$ \\
        ~ & $n_{\vert\mathcal{S}\vert+1}^{(1)}$ & $n_{\vert\mathcal{S}\vert+1}^{(2)}$ & $\ldots$ & $n_{\vert\mathcal{S}\vert+1}^{(\iunit)}$ & $\ldots$ & $n_{\vert\mathcal{S}\vert+1}^{(\nunit)}$ \\
        \bottomrule
    \end{tabular}
\end{table}
}

\begin{figure*}
    \centering
    \includegraphics[width=\textwidth]{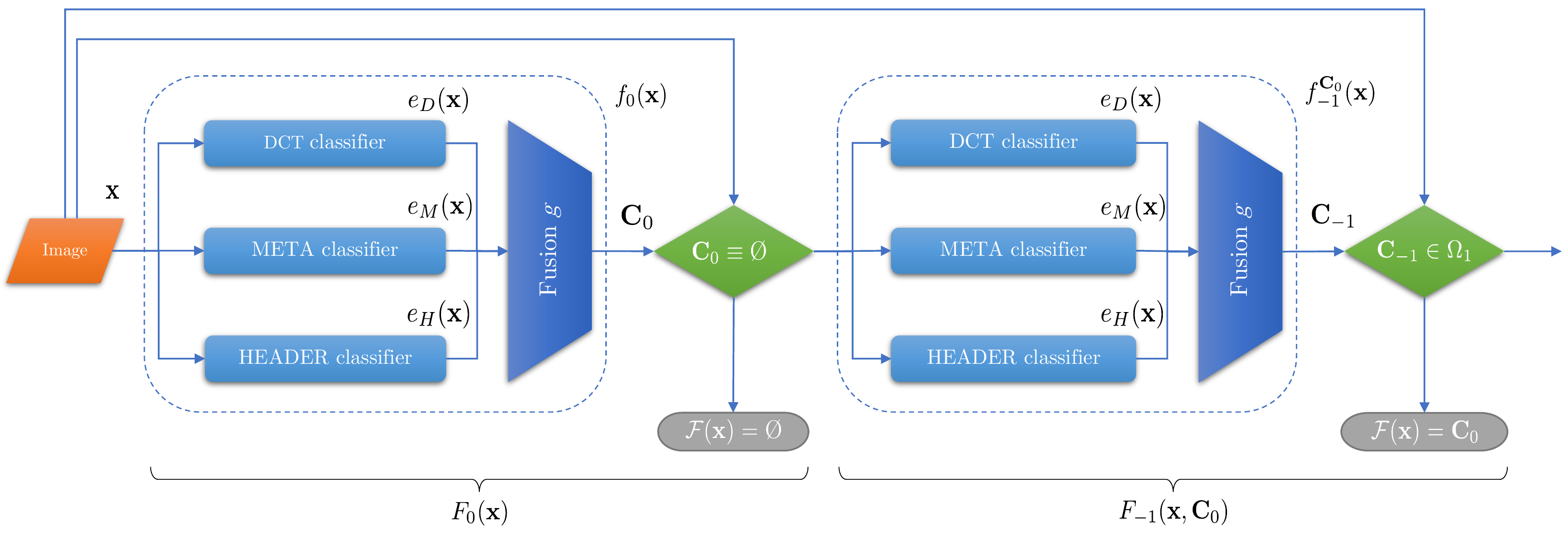}
    \caption{Framework architecture for multi-clue reconstruction of sharing chains. The input image $\x$ is fed into a cascade of backtracking blocks, $F_{-\ell}$, each one dedicated to identifying chains of length up to $\ell+1$. Every step consists of an ensemble of parallel classifiers, which analyse $\x$ by also taking into account the information from previous steps, and a fusion module that provides the overall decision. Stopping conditions intervene when $F_{-\ell}$ receives a chain of length $\ell-1$, i.e., the end of the chain has been reached. }
    \label{fig:cascade}
\end{figure*}

Let us assume that the input $\x$ is mapped into a set of decisions $e_k(\x)$, $k=1,\ldots,K$, that corresponds to the focal unit $u_\iunit$ of the BKS. From the lookup table, we can derive the posterior probability of associating input $\x$ to a label $y$ given the focal unit $u_\iunit$ as follows
\begin{equation}
    P\left(f(\x)=y \mid u_\iunit\right) = \frac{n_y^{(\iunit)}}{\sum_{j=1}^{\vert\mathcal{S}\vert + 1} n_j^{(\iunit)}}
\end{equation}

The optimal combined decision is therefore
\begin{equation}
    \hat{y}=\underset{y}{\operatorname{argmax}} \ P\left(f(\x)=y\mid u_\iunit\right)
\end{equation}

In BKS fusion, there is also the possibility of an input $\x$ being \emph{rejected}, meaning that while the individual decisions are valid their combination is impossible. In practice, a rejection occurs when the focal unit is empty, i.e., no training samples have been mapped into $u_\iunit$, or when the maximum value of the estimated distribution $n_y^{(\iunit)}$ is non-unique. 

In the cascade architecture, a rejection from the fusion module stops the reconstruction process seamlessly, since its effect is equivalent to the case when the end-point of a sharing chain is reached (see Equation~\ref{eq:block}).

Figure~\ref{fig:cascade} depicts a comprehensive representation of the cascaded architecture, exemplified by two backtracking blocks, including the fusion modules and the stopping conditions; the classifiers related to the three adopted descriptors (DCT, META, HEADER) are denoted by $e_D$, $e_M$, $e_H$.

\section{Experiments and Results}\label{sec:results}

% *** SETTING ***

\subsection{Experimental setting}

The presented framework is highly flexible, being adaptable to a different number $K$ of feature classifiers, set $\mathcal{S}$ of sharing platforms, and length $L$ of the longest detectable sharing chain. 

In our experiments, we implemented the system with the following configuration:
\begin{itemize}
    \item $K=3$ feature descriptors (\DCT, \META, \HEADER); %, and thus each function $f^\C_{-\ell}$ combines three classifiers, denoted by $e_D$, $e_M$, $e_H$ (Figure~\ref{fig:cascade});
    \item $\mathcal{S}=\{\lab{FB}{}{},\lab{FL}{}{},\lab{TW}{}{}\}$, where the labels denote Facebook, Flickr and Twitter, respectively;
    \item $L=3$, therefore the system $\F_L$ is composed of three backtracking blocks, $F_0$, $F_{-1}$, $F_{-2}$.
\end{itemize}

It follows that the set of all possible sharing chains identified by the system is $\Omega_3=\mathcal{S}\cup\mathcal{S}^2\cup\mathcal{S}^3$, where:
\begin{itemize}
    \item $\mathcal{S}^2=\{\lab{FB}{FB}{},\lab{FL}{FB}{},\lab{TW}{FB}{},\ldots\},\quad\vert\mathcal{S}^2\vert=9$;
    \item $\mathcal{S}^3=\{\lab{FB}{FB}{FB},\lab{FL}{FB}{FB},\ldots\},\quad\vert\mathcal{S}^3\vert=27$;
\end{itemize}
and thus $\vert\Omega_3\vert=39$.

\smallskip

\subsubsection{Dataset description}
the framework was trained and tested on the \textbf{R-SMUD} dataset \cite{phan2019}, containing images shared on the three social platforms in $\mathcal{S}$. Images are generated starting from the RAISE dataset \cite{dang2015raise} by cropping 50 raw format images on the top-left corner, with fixed aspect ratio of 9:16 and different resolutions: 377x600, 1012x1800 and 1687x3000. Each of the obtained crops was then compressed with six quality factors (50, 60, 70, 80, 90, 100) and then shared up to a maximum of 3 times through different combinations of the three platforms in $\mathcal{S}$, providing a total of 35100 images and $\vert\Omega_3\vert=39$ classes. The dataset was then divided into training, validation and test subsets, with a 60/20/20 split. 

\smallskip

\subsubsection{Classifiers and training}
as described in Section~\ref{ssec:fusion}, each backtracking block contains an ensemble of trained classifiers, one for each combination of feature descriptor and output of the previous block.

Given the number $K$ of feature descriptors, and given that one detector $f^\C_{-\ell}$ is needed for each partially reconstructed chain $\C \in \Omega_\ell \setminus \Omega_{\ell-1}$ (see Section~\ref{ssec:cascade}), we can derive the number of classifiers needed at each backtracking block:
\begin{itemize}
    \item $F_0 \longrightarrow K=3$
    \item $F_{-1} \longrightarrow K\cdot\vert\Omega_1\vert=9$
    \item $F_{-2} \longrightarrow K\cdot\vert\Omega_2\setminus\Omega_1\vert=27$
\end{itemize}

We also recall from Section~\ref{ssec:fusion} that the number of output classes of each detector, regardless of the step index $\ell$ and the partial chain $\C$, is equal to $\vert\mathcal{S}\vert+1=4$. For instance, if $F_0(\x)=\C_0=\lab{FB}{}{}$, then the output of $F_{-1}(\x,\C_0)$ can either be $\C_{-1}=\lab{FB}{}{}$ or $\C_{-1}=\lab{*}{FB}{}\in\mathcal{B}(\C_0)$, where $*$ indicates whatever element in $\mathcal{S}$.

Each detector was implemented as a Random Forest Classifier (RFC), with fixed hyper-parameters. In particular, we used a number of estimators equal to 10, balancing accuracy and model complexity. 

Classifiers were trained on the training subset of R-SMUD, containing 21060 images equally distributed among the 39 classes. 
Detectors belonging to $F_0$ and $F_{-1}$, which are dedicated to the classification of shorter chains than $F_{-2}$, were trained after re-mapping the original 39 classes as follows:
\begin{itemize}
    \item $F_0$ is trained on chains in $\Omega_1$, $$\left(\C[-2],\ \C[-1],\ \C[0]\right)\in\Omega_3 \longrightarrow \C[0]\in\Omega_1;$$
    \item $F_{-1}$ is trained on chains in $\Omega_2$, $$\left(\C[-2],\ \C[-1],\ \C[0]\right)\in\Omega_3 \longrightarrow \left(\C[-1],\ \C[0]\right)\in\Omega_2.$$
\end{itemize}

Such training strategy allows fully exploiting the amount of samples in the dataset and at the same time recreating a more realistic scenario: as an example, the detection of $\C[0]$ at step $F_0$ is carried out among chains of different lengths, having $\C[0]$ as their last step.

Since steps $F_{-\ell}$, $\ell>0$, discriminate chains that are known up to $\C[-(\ell-1)]$, the related classifiers need to be trained on smaller subsets of samples, obtained by fixing $\C[0], \C[-1], \ldots, 
\C[-(\ell-1)]$. In general, the fraction of training samples available to train step $F_{-\ell}$ is $\nicefrac{1}{\vert\Omega_\ell\setminus\Omega_{\ell-1}\vert}$. For instance, at step $F_{-1}$ we have to split the training according to the possible outputs of $F_0$, and thus we get $\nicefrac{1}{\vert\Omega_1\vert}=\nicefrac{1}{3}$, corresponding to 7020 images; similarly, at step $F_{-2}$ we get $\nicefrac{1}{\vert\Omega_2\setminus\Omega_1\vert}=\nicefrac{1}{9}$, corresponding to 2340 images.

As for the fusion part, the number of BKS modules required at each backtracking block equals the number of classes output by the related previous block, similarly to the classifiers. Each fusion module was trained on the validation subset of R-SMUD (7020 images) to prevent biasing the interaction between classifiers and fusions (Section~\ref{ssec:fusion}). The same dataset partitioning described above for steps $F_{-\ell}$, $\ell>0$, was carried out for the training of fusion modules.

Finally, the end-to-end system was evaluated on the test set, containing 7020 images. The experimental results are reported and commented in the next paragraphs.

% *** RESULTS ***

\subsection{Reconstruction results}

Each step of the cascade architecture was evaluated separately, in order to observe the reconstruction behavior throughout the pipeline. In detail, we ran the system on the test set of R-SMUD and measured the detection accuracy at the output of each backtracking block, $F_{-\ell}$, $\ell=0,1,2$. We recall that the number of output classes at each block is equal to $\vert\Omega_{\ell+1}\vert$, hence: 3 classes at $F_0$; 12 classes at $F_{-1}$; 39 classes at $F_{-2}$.

Additionally, we ran the experiments by including different subsets of feature descriptors in the system, to better assess the individual contributions: first, we tested {\DCT}, {\META} and {\HEADER} features by themselves; then, the fusion of pairs of features; and finally, the fusion of all the three.

\setlength{\tabcolsep}{4.5pt}
\begin{table}[t]
    \centering
    \caption{Per-step performance of the cascade system.}
    \begin{tabular}{rcccccc}
         \toprule    
         & \multicolumn{2}{c}{$F_0$} & \multicolumn{2}{c}{$F_{-1}$} & \multicolumn{2}{c}{$F_{-2}$}\\
         & \multicolumn{2}{c}{3 classes} & \multicolumn{2}{c}{12 classes} & \multicolumn{2}{c}{39 classes}\\
         \midrule
        {\scriptsize Single classifiers} & ACC &  & ACC &  & ACC & \\
         \midrule
         {\scriptsize {\DCT}} & 0.8634 &  & 0.4620 & & 0.1849 & \\
         {\scriptsize {\META}} & 0.9296 &  & 0.5350 & & 0.2959 &\\
         {\scriptsize {\HEADER}} & \textbf{1.0000} &  & 0.5994 & & 0.2289 &\\
         {\scriptsize Random guess} & 0.3333 & & 0.0833 & & 0.0256 & \\
         \midrule
         {\scriptsize Fused classifiers} & ACC & REJ & ACC & REJ & ACC & REJ\\
         \midrule
         {\tiny {\DCT}+{\META}} & 0.9296 & 0.0000 & 0.6096 & 0.0198 & 0.3475 & 0.1849\\
         {\tiny {\META}+{\HEADER}} & \textbf{1.0000} & 0.0000 & 0.7934 & 0.0188 & \textbf{0.5576} & 0.2325\\
         {\tiny {\DCT}+{\META}+{\HEADER}} & \textbf{1.0000} & 0.0000 & \textbf{0.7981} & 0.0208 & 0.5465 & 0.1949\\
         \bottomrule
    \end{tabular}
    \label{tab:cascade}
\end{table}

Table~\ref{tab:cascade} reports an overview of accuracy values for each step of the cascade and for all feature configurations; for fused classifiers, we also report the rejection rates, i.e., the percentage of test samples rejected by the BKS fusion. At a macroscopic level, we can observe the following: (i) fused classifiers perform better than (or on par with) single ones, suggesting that the usage of heterogeneous feature descriptors is beneficial for platform provenance analysis; (ii) {\HEADER} features allow a \emph{deterministic} identification of the last sharing platform, which is preserved even when fused with other classifiers; (iii) the overall accuracy values tend to decrease as we proceed through the cascade, while rejection rates increase. 

\begin{figure}[t]
    \centering
    \subfloat[{\tiny {\DCT}, $F_0$.}]{\includegraphics[width=0.33\columnwidth]{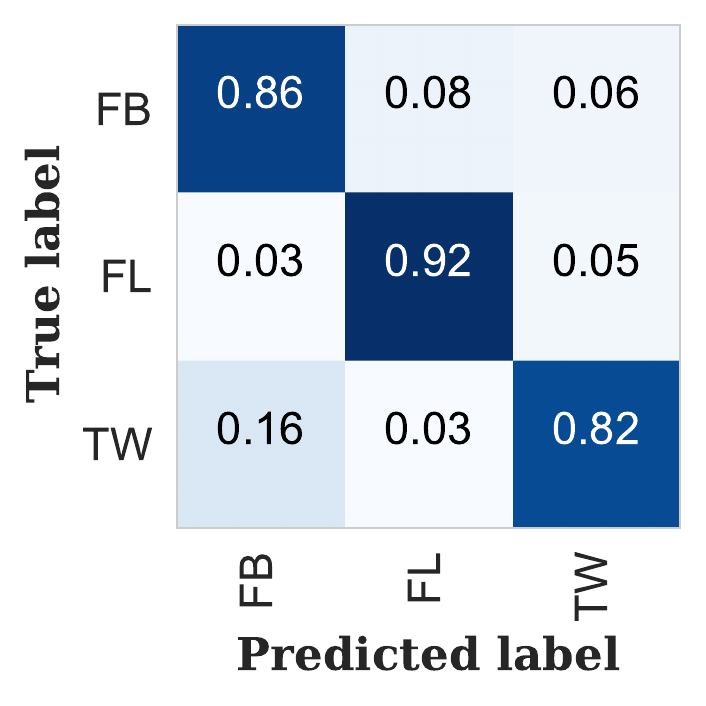}}
    \subfloat[{\tiny {\META}, $F_0$.}]{\includegraphics[width=0.33\columnwidth]{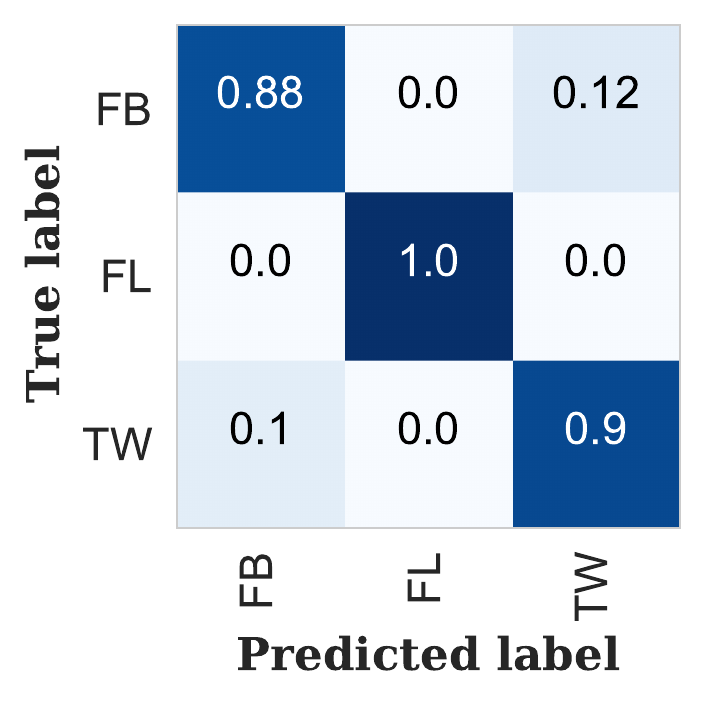}}
    \subfloat[{\tiny {\HEADER}, $F_0$.}]{\includegraphics[width=0.33\columnwidth]{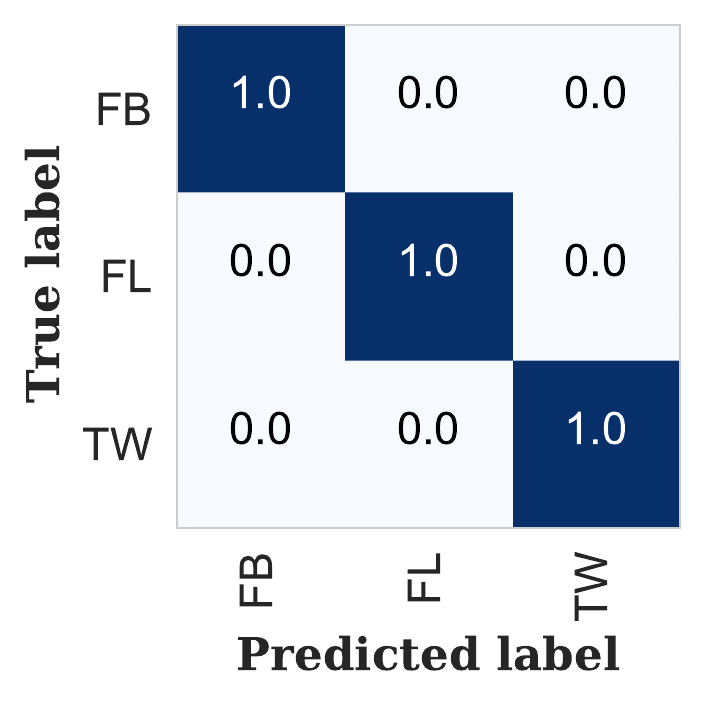}}\\
    \subfloat[{\tiny {\DCT}+{\META}, $F_0$.}]{\includegraphics[width=0.33\columnwidth]{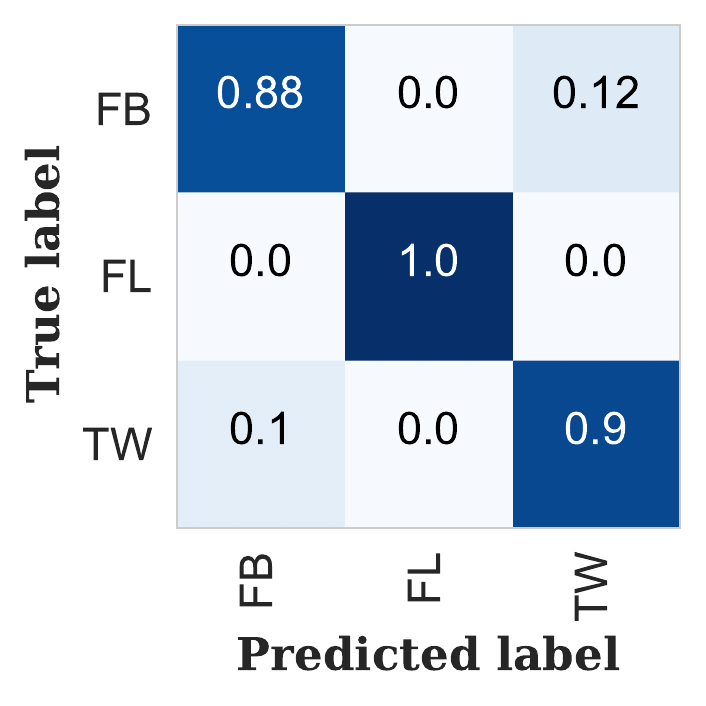}}
    \subfloat[{\tiny {\META}+{\HEADER}, $F_0$.}]{\includegraphics[width=0.33\columnwidth]{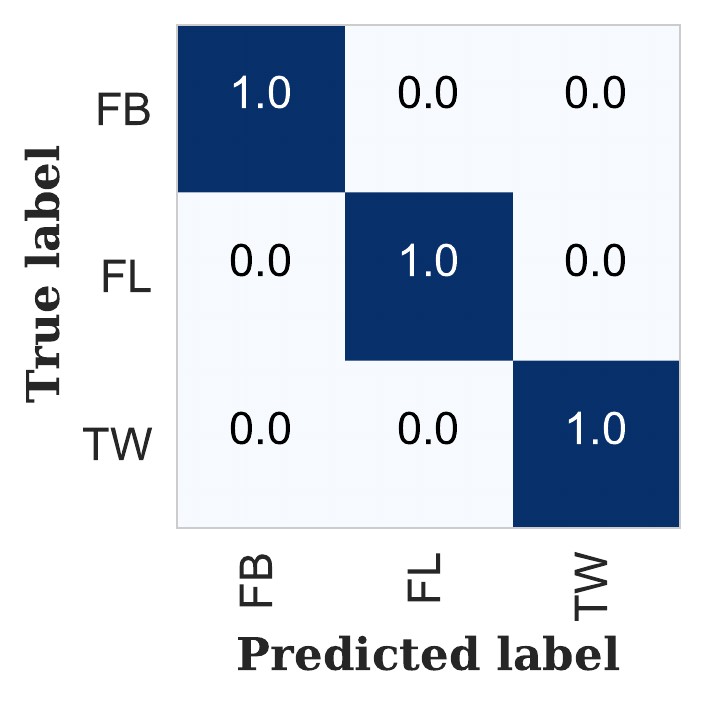}}
    \subfloat[{\tiny {\DCT}+{\META}+{\HEADER}, $F_0$.}]{\includegraphics[width=0.33\columnwidth]{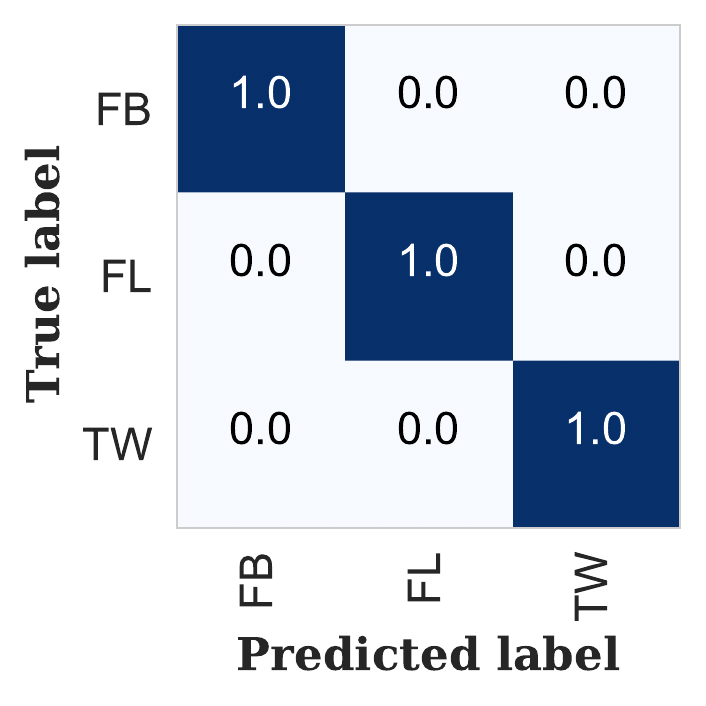}}
    \caption{Detection performance of backtracking block $F_0$, which identifies the last platform in the sharing chain, $\C[0]$; tests carried out with individual feature sets (a--c), pair fusions (d--e) and triplet fusion (f).}
    \label{fig:step0}
\end{figure}

\begin{figure*}[t]
    \centering
    \subfloat[{\DCT}, $F_{-1}$.]{\includegraphics[width=0.33\textwidth]{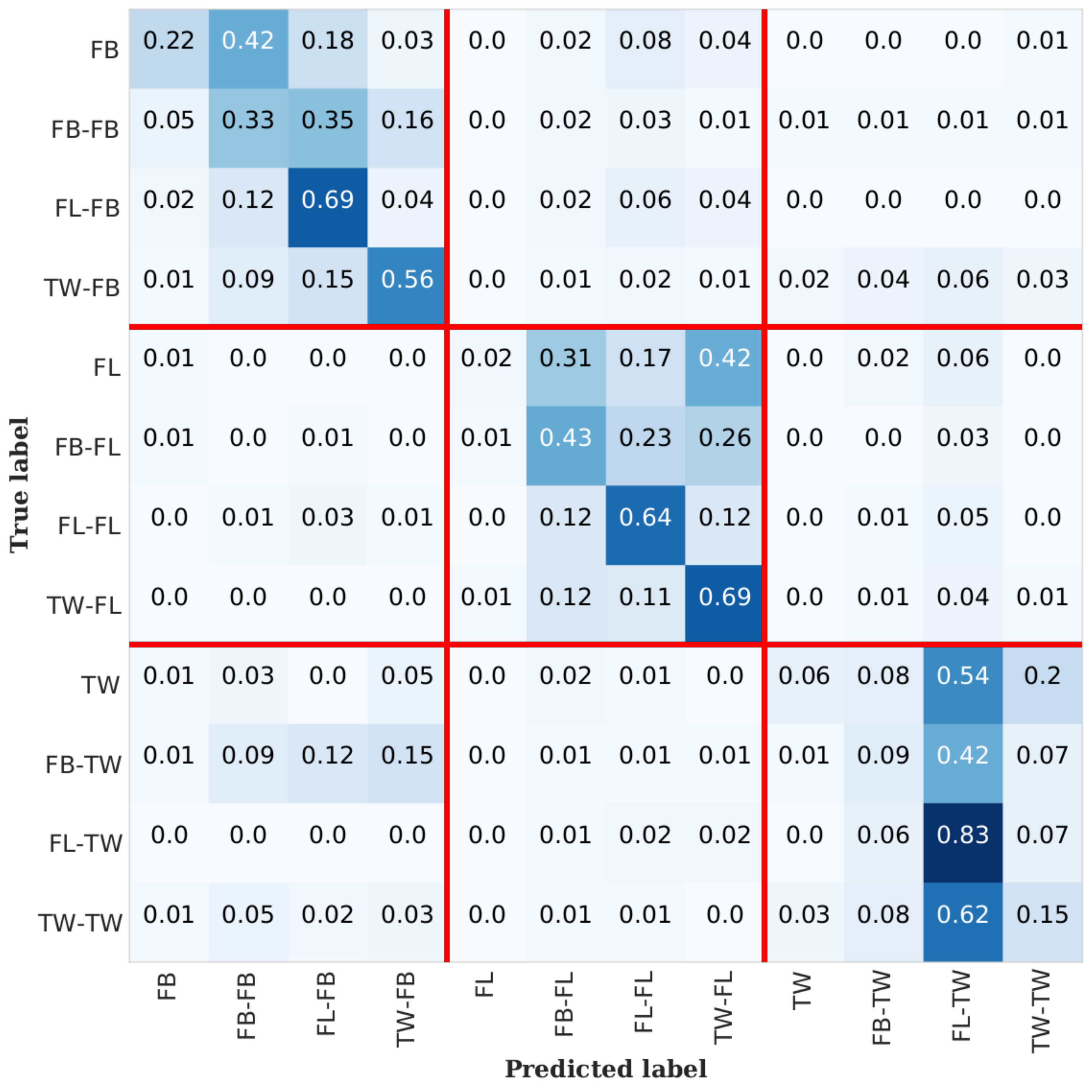}}
    \subfloat[{\META}, $F_{-1}$.]{\includegraphics[width=0.33\textwidth]{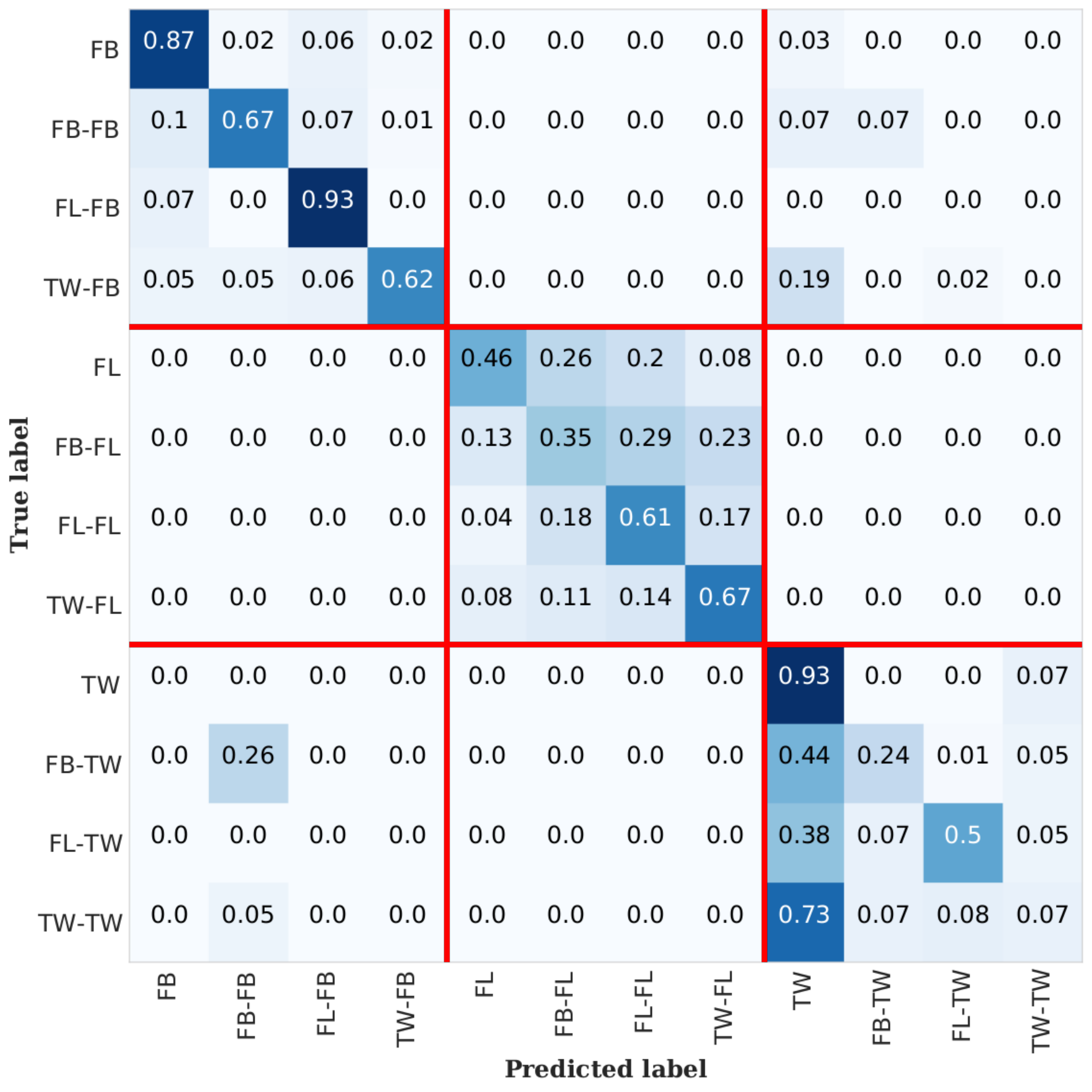}}
    \subfloat[{\HEADER}, $F_{-1}$.]{\includegraphics[width=0.33\textwidth]{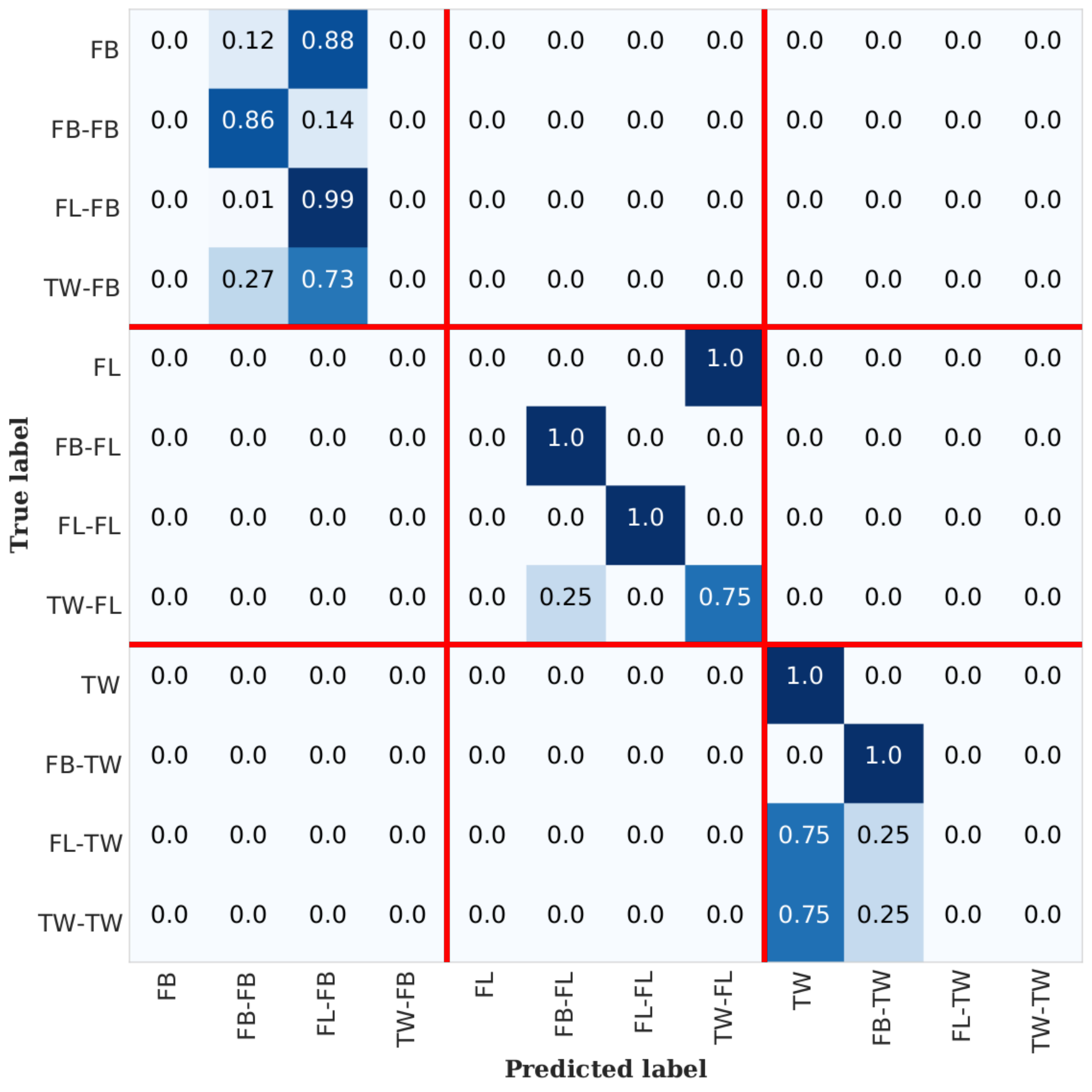}}\\
    \subfloat[{\DCT}+{\META}, $F_{-1}$.]{\includegraphics[width=0.33\textwidth]{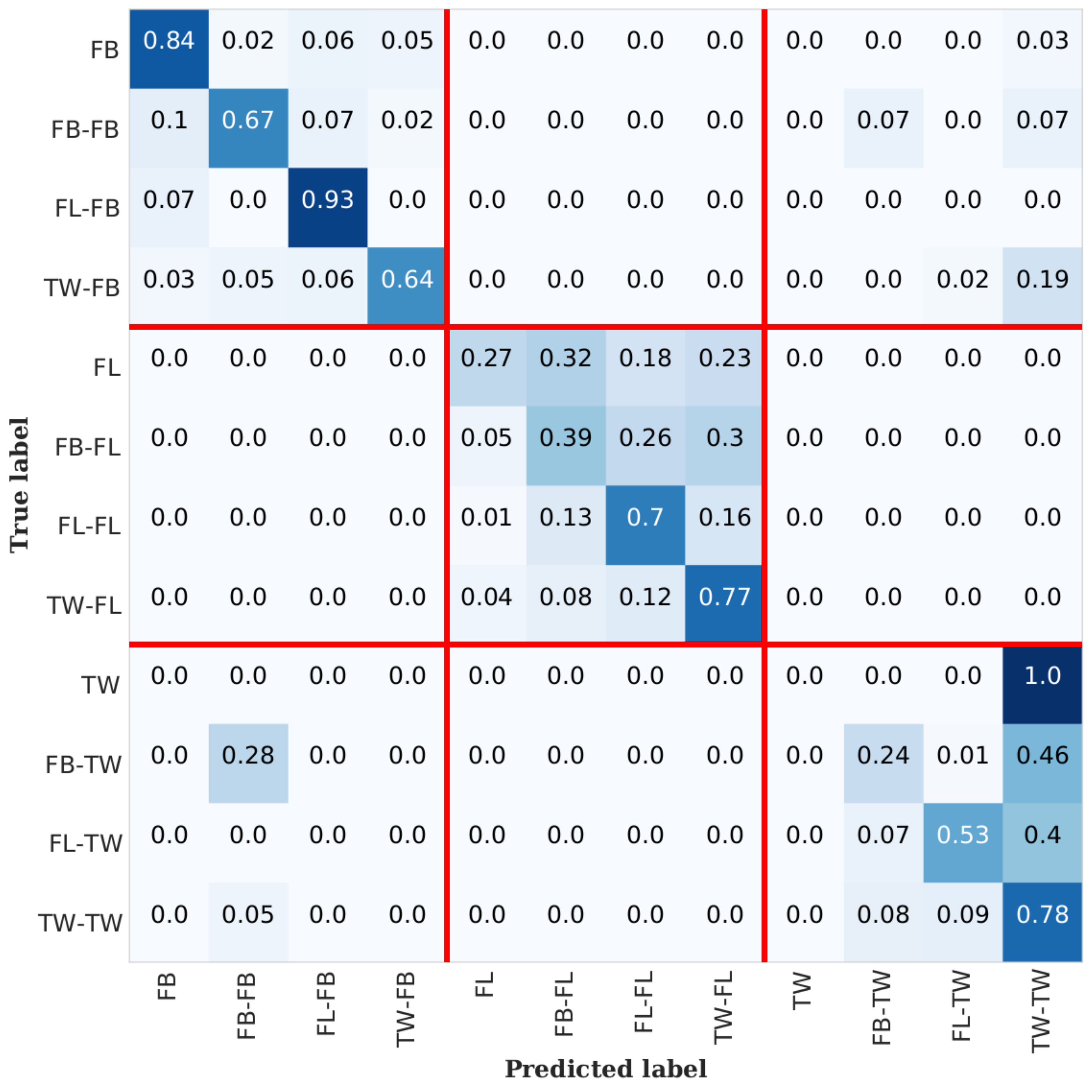}}
    \subfloat[{\META}+{\HEADER}, $F_{-1}$.]{\includegraphics[width=0.33\textwidth]{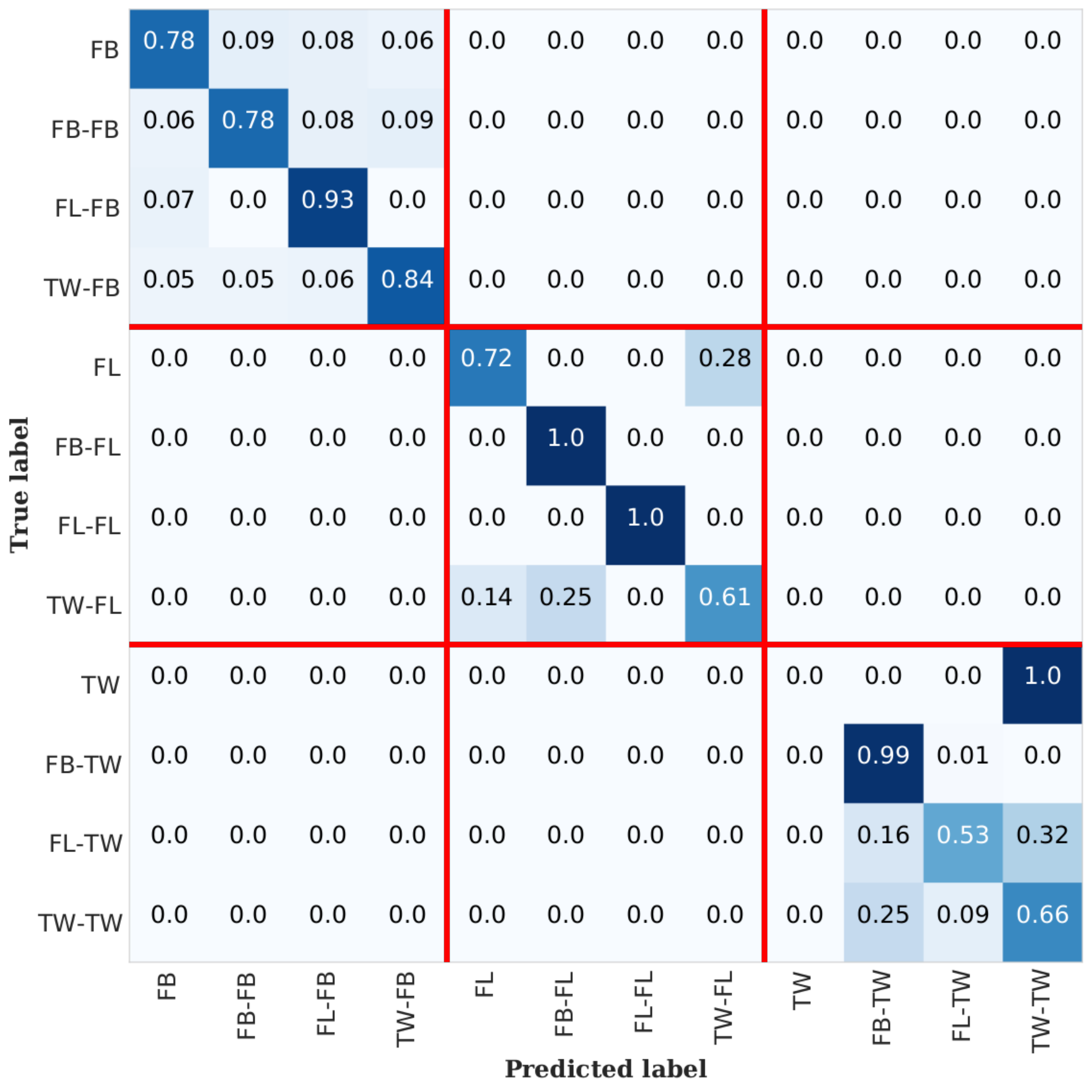}}
    \subfloat[{\DCT}+{\META}+{\HEADER}, $F_{-1}$.]{\includegraphics[width=0.33\textwidth]{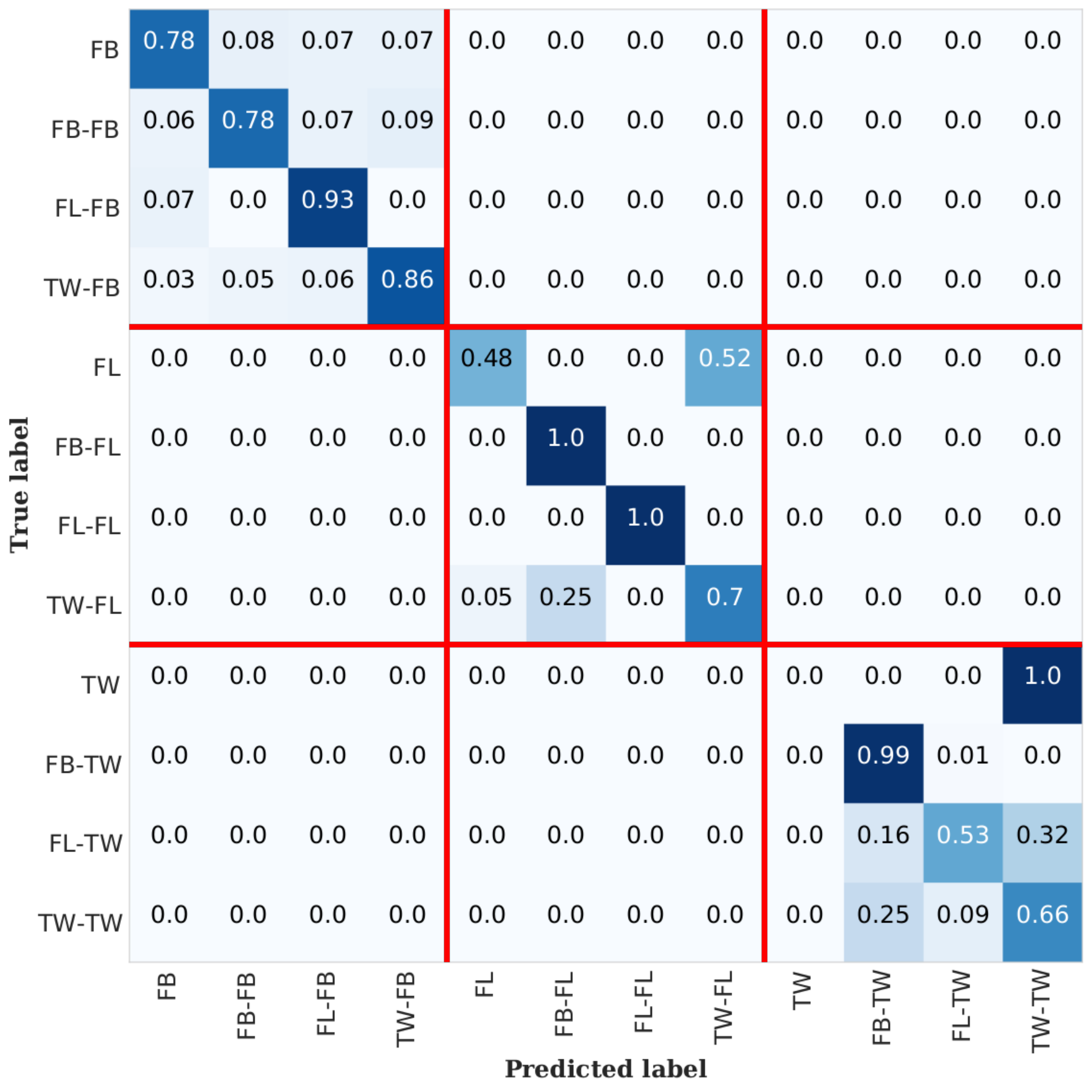}}
    \caption{Detection performance of backtracking block $F_{-1}$, which identifies the last two platforms in the sharing chain, ($\C[-1],\C[0]$); tests carried out with individual feature sets (a--c), pair fusions (d--e) and triplet fusion (f). Red squares highlight the subsets of classes having $\C[0]$ in common; note how most confusions are confined within the diagonal squares (all of them, when {\HEADER} is used).}
    \label{fig:step1}
\end{figure*}

To deeper investigate these preliminary observations, the following paragraphs illustrate in details, with the aid of confusion matrices, the performance at each step of the cascade.

\smallskip

% *** F0 *** 

\subsubsection{$F_0$ step} 

Figure~\ref{fig:step0} shows the 3-by-3 confusion matrices obtained at step $F_{0}$ with different feature configurations.

The last sharing step is unsurprisingly the easiest one to be identified, since the forensic traces related to that platform are still intact, while they tend to vanish or blend as we move backwards along the sharing chain. In fact, all standalone detectors are able to reach high accuracy values by themselves (Figure~\ref{fig:step0}a--c). Nevertheless, one can observe some interesting results. First, {\HEADER} features are able to detect the last step $\C[0]$ with deterministic accuracy; in fact, the presence and frequency of JPEG header markers are closely related to the processing carried out by the last sharing platform during upload. Secondly, we can note that BKS preserves the performance of the best classifiers among those included in the fusion: when {\DCT} and {\META} descriptors are combined, the results are identical to the ones obtained with {\META} alone (Figure~\ref{fig:step0}d); when {\HEADER} is present in the fusion, instead, perfect detection is preserved (Figure~\ref{fig:step0}e--f). This characteristic of {\HEADER} features is key at the first backtracking block of the cascade, as it guarantees to precisely identify the last step of the chain, thus avoiding the propagation of errors throughout the pipeline.

\begin{figure*}[t]
    \centering
    \subfloat[{$\C[0]=\ $}FB]{\includegraphics[width=0.33\textwidth]{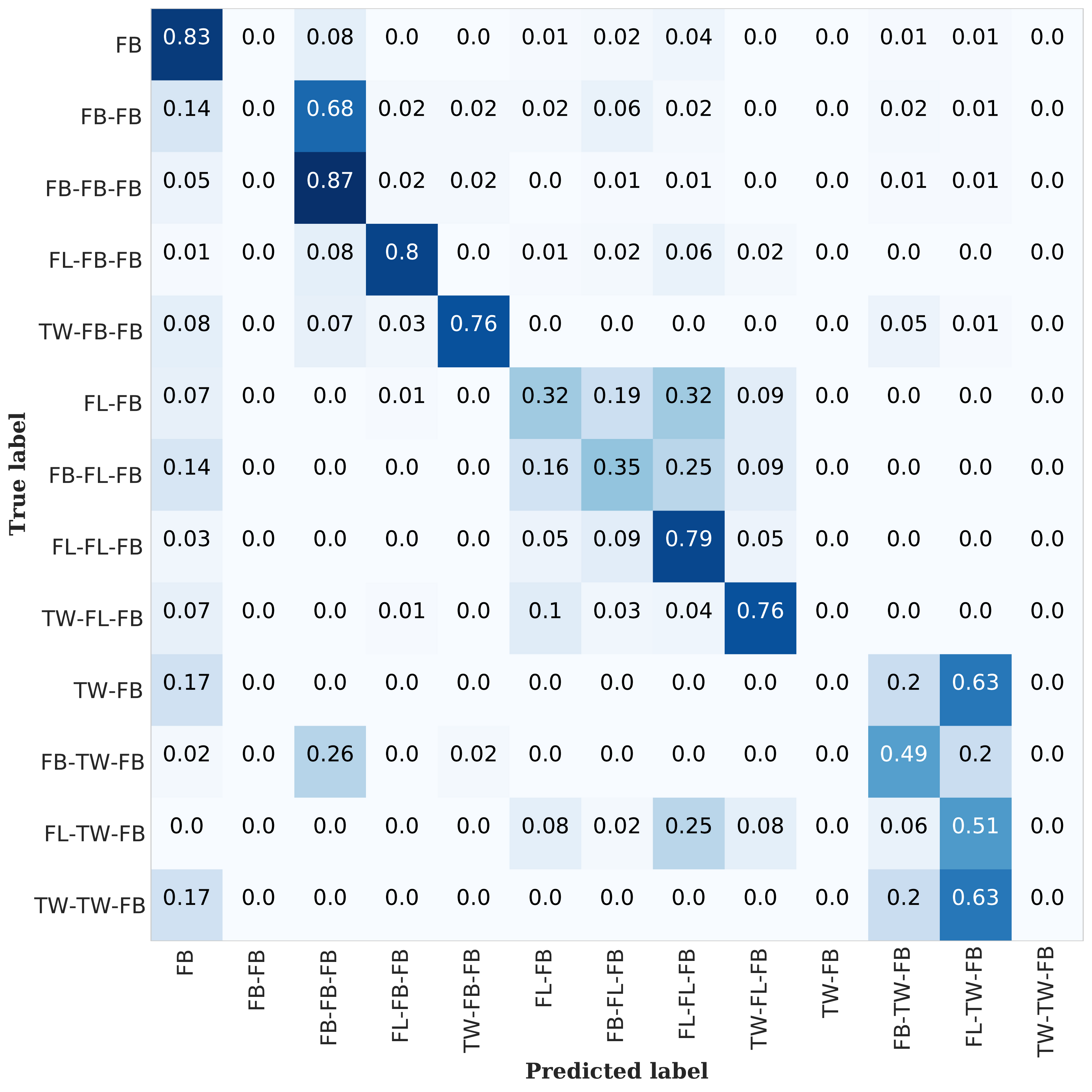}}
    \subfloat[{$\C[0]=\ $}FL]{\includegraphics[width=0.33\textwidth]{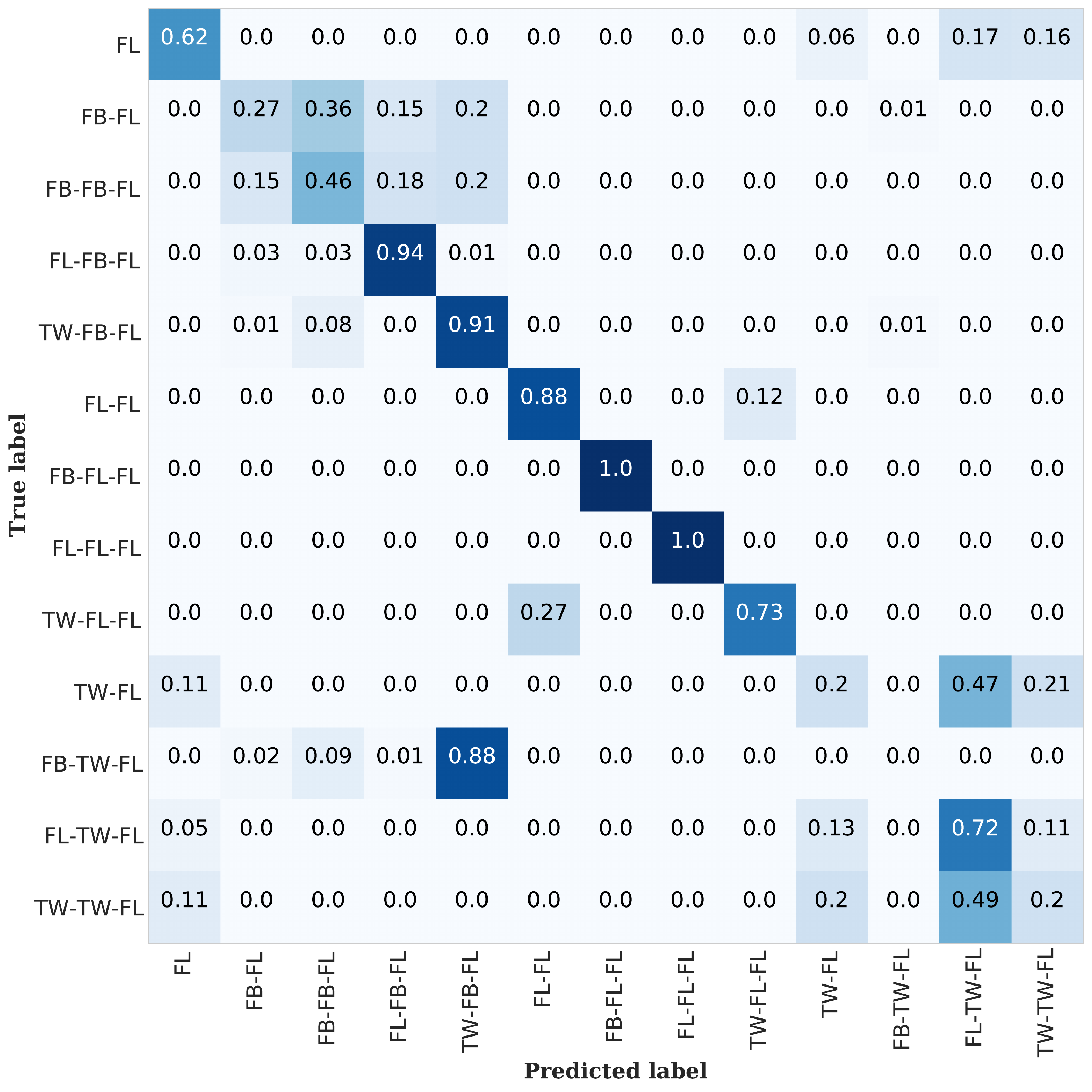}}
    \subfloat[{$\C[0]=\ $}TW]{\includegraphics[width=0.33\textwidth]{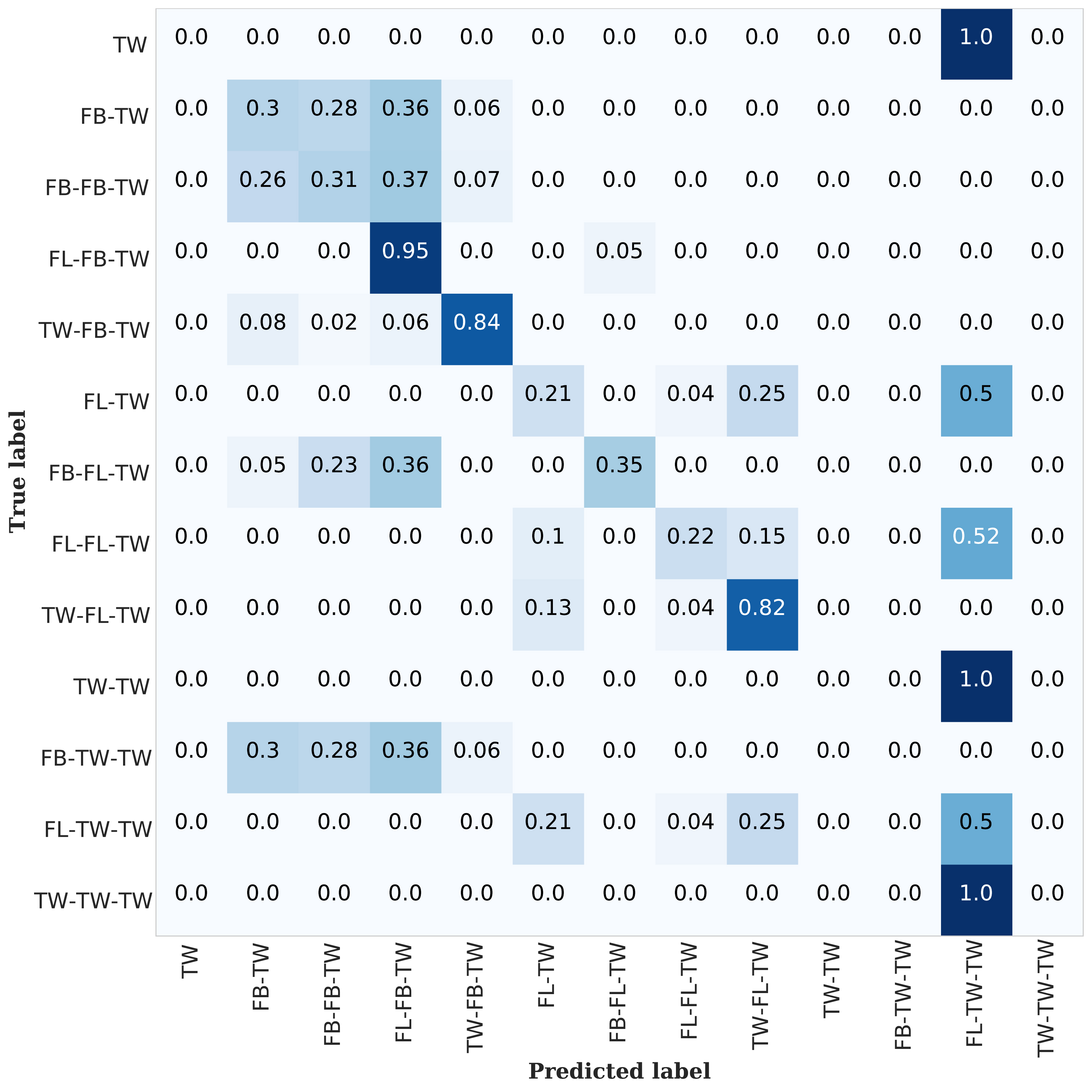}}
    \caption{Detection performance of backtracking block $F_{-2}$, which identifies the last three platforms in the sharing chain, ($\C[-2],\C[-1],\C[0]$) and constitutes the final output of the implemented system $\F$. The test is carried out with the fusion of all three feature descriptors (see Table~\ref{tab:cascade} for the results with different feature configurations). The overall 39-by-39 confusion matrix is reported by means of the three diagonal blocks related to chains with $\C[0]$ in common, namely Facebook (a), Flickr (b) and Twitter (c); all elements outside the diagonal blocks are empty, meaning that the last sharing step is always detected correctly.}
    \label{fig:step2}
\end{figure*}

\smallskip

% *** F-1 ***

\subsubsection{$F_{-1}$ step}

Figure~\ref{fig:step1} shows the 12-by-12 confusion matrices obtained at step $F_{-1}$ with different feature configurations; the red grid in overlay is meant to highlight the sub-squares related to chains having $\C[0]$ in common. 

The reconstruction of $\C[-1]$ allows observing additional interesting results on the behavior of the cascaded system. The results obtained with the individual feature sets (Figure~\ref{fig:step1}a-c) show that classification errors mostly occur among sharing chains having the same platform as last step. For instance, by looking at the top-left 4-by-4 square of the DCT confusion matrix (Figure~\ref{fig:step1}a), which is related to the chains with $\C[0]=\lab{FB}{}{}$, we can observe how confusions are mostly confined within it; the same result appears in the other sub-squares and for all descriptors. This is a direct consequence of the cascaded approach, which is preserving the performance of the first step throughout the rest of the pipeline. In particular, we can re-observe the perfect classification of $\C[0]$ obtained with {\HEADER} features: in Figure~\ref{fig:step1}c, 100\% of classifications is confined within the three respective 4-by-4 squares; however, {\HEADER} alone provides a poor classification of $\C[-1]$, specially for chains ending in $\lab{FB}{}{}$ and $\lab{TW}{}{}$ (top-left and bottom-right squares). Nevertheless, the fusion function is able to compensate this problem by exploiting the information from the other classifiers: as we can see in Figure~\ref{fig:step1}f and in Table~\ref{tab:cascade}, the fusion of all the three descriptors allow reaching an overall accuracy close to 80\%, with a rejection rate of only 2\%. Also, we observe that most of the errors occurring in fused configurations fall in the bottom-right square, which is related to sharing chains with $\C[0]=\lab{TW}{}{}$ (more on this in Section~\ref{sec:separability}).

\smallskip

% *** F-2 ***

\subsubsection{$F_{-2}$ step} 

Figure~\ref{fig:step2} shows one of the 39-by-39 confusion matrices obtained at step $F_{-2}$; due to the considerable size of the matrices at this step, we only report the one related to the fusion of all three classifiers; also, since the matrix is perfectly empty outside the three main blocks on the diagonal, which contain chains having $\C[0]$ in common, we just report said blocks separately, in Figure~\ref{fig:step2}a--c.

At the last step of the pipeline, the final classification involves sharing chains of any length up to $L=3$, which corresponds to $\vert\Omega_3\vert=39$ classes. Despite the high number of classes, the system is able to reach a 55\% overall accuracy (random guess is 2.56\%) with the fusion of all feature descriptors. However, we also observe a dramatic increase in the rejection rate with respect to the previous steps (see Table~\ref{tab:cascade}). Moreover, Figure~\ref{fig:step2}c highlights a clear performance drop in the TW-related square, with respect to the other two, suggesting that sharing chains with $\C[0]=\lab{TW}{}{}$ are somehow more difficult to distinguish.
To explain these results, we first analysed the per-class rejection distribution, discovering that 100\% of rejections occurred in sharing chains having $\lab{TW}{}{}$ as their last step, which is also in agreement with the performance drop in Figure~\ref{fig:step2}c. 

Such initial clues on the difficulties introduced by the presence of Twitter in sharing chains motivated us to conduct a deeper analysis on feature separability, which is the topic of Section~\ref{sec:separability}. 

\smallskip

% *** SOTA COMPARISON ***

\subsubsection{State-of-the-art comparison}\label{sec:sota}

Table~\ref{tab:sota} reports a comparison of state-of-the-art methods for the image recycling problem. Solutions in~\cite{caldelli2017,amerini2017} employ histograms of {\DCT} coefficients combined with a deep learning approach based on convolutional neural networks (CNNs). In~\cite{phan2019} the authors propose a patch-based CNN in two different configurations: the first one receives only {\DCT} features in input (P-CNN), while the second one operates a feature fusion of {\DCT} coefficients and metadata (P-CNN-FF). For the proposed method, we report the results related to the fusion of {\DCT}, {\META} and {\HEADER} features. All results are obtained on the R-SMUD dataset~\cite{phan2019} and reported separately for chains of up to one ($\C_0$), two ($\C_{-1}$) and three ($\C_{-2}$) sharing steps.

\setlength{\tabcolsep}{6pt}
\begin{table}[ht]
    \centering
    \caption{Method comparison for image recycling.}
    \label{tab:sota}
    \begin{tabular}{cccc}
    \toprule
    ~ & \multicolumn{3}{c}{Accuracy on {\scriptsize R-SMUD \cite{phan2019}}} \\
    ~ & $\C_0$ & $\C_{-1}$ & $\C_{-2}$ \\
    Method & 3 classes & 12 classes & 39 classes \\
    \midrule
    \cite{caldelli2017} & 0.9370 & 0.3991 & 0.1729 \\
    \cite{amerini2017} & 0.9481 & 0.4518 & 0.1695 \\
    P-CNN \cite{phan2019} & 0.8963 & 0.4324 & 0.1932 \\
    P-CNN-FF \cite{phan2019} & 0.9987 & 0.6591 & 0.3618 \\
    Proposed & \textbf{1.0000} & \textbf{0.7981} & \textbf{0.5465} \\
    \bottomrule
    \end{tabular}
\end{table}

% *** SEPARABILITY ***

\subsection{Separability analysis}\label{sec:separability}

%In order to explain the results in the previous section, we studied the level of separability of the different sharing chains. First of all, we noted that images shared with similar chains (e.g., $\lab{FB}{FB}{}$ and $\lab{FB}{FB}{FB}$) and in particular those that were at some point shared on Twitter present identical feature vectors.

\begin{figure*}[ht]
    \centering
    \subfloat[Average per-class LSR computed for {\DCT}, {\META} and {\HEADER} features.]{\includegraphics[width=\textwidth]{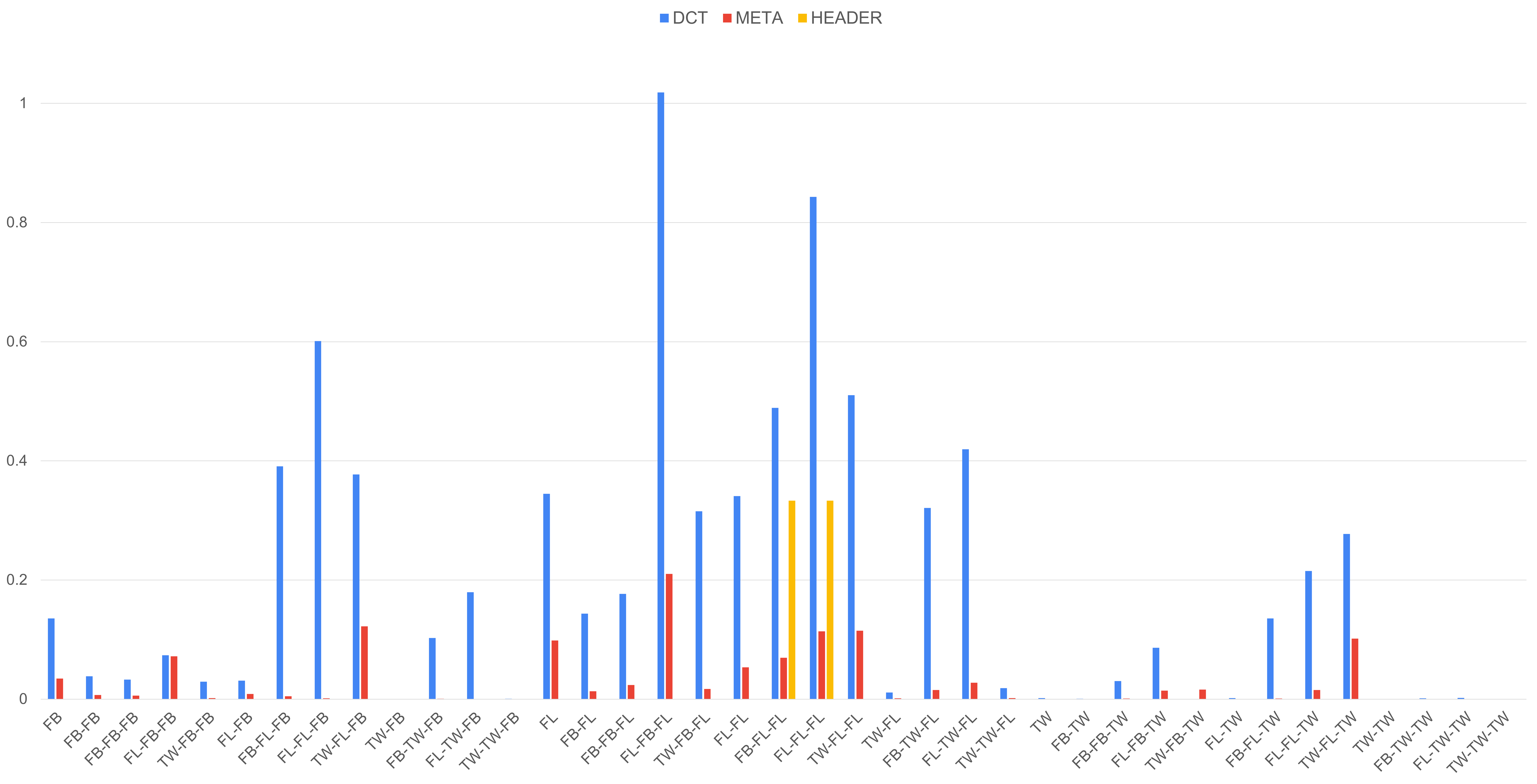}}\\
    \subfloat[Aggregated LSR for FB.]{\includegraphics[width=0.32\textwidth]{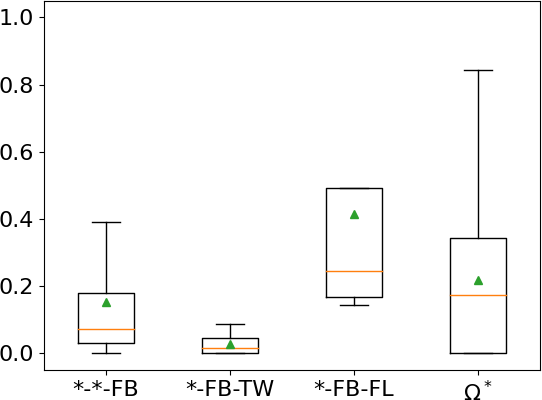}}\quad
    \subfloat[Aggregated LSR for FL.]{\includegraphics[width=0.32\textwidth]{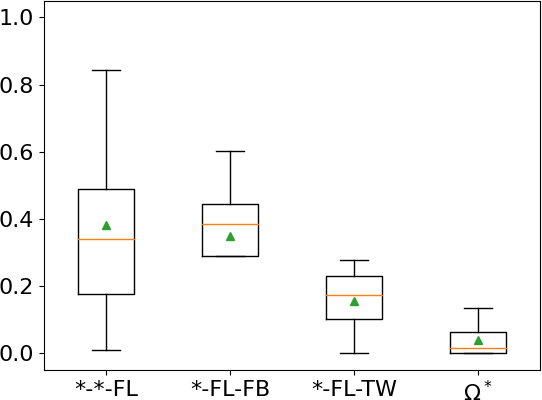}}\quad
    \subfloat[Aggregated LSR for TW.]{\includegraphics[width=0.32\textwidth]{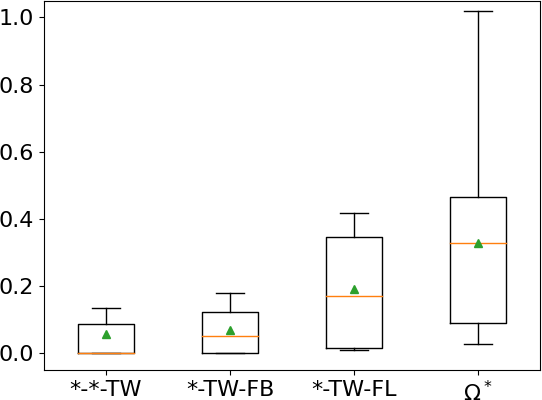}}
    \caption{Average Local Set Radius (LSR) for each sharing chain and for each of the three feature sets, namely {\DCT}, {\META} and {\HEADER} (a), and boxplots showing LSR values computed for {\DCT} features and aggregated by platform (b--d); each boxplot represents, from left to right, the chains ending with a specific sharing platform, the two possible sets of chains having that same platform in $\C[-1]$, and all the other classes ($\Omega^*$). The boxes represent the lower and upper quartile, with whisker showing the minumum and maximum values, the median represented as a yellow line and the mean as a green triangle.  Low LSR values are associated to chains ending in or containing Twitter, which makes them hardly separable in the feature space.}
    \label{fig:LSR}
\end{figure*}

To formally study feature separability, we started from the definition of the ratio of intra/extra-class nearest-neighbor distance~\cite{ho2002data}, which is formulated as
\begin{equation}
    IER = \frac{\sum_{i=1}^{n} d(\x_i, NN(\x_i)  \in \j_i)}{\sum_{i=1}^{n} d(\x_i, NN(\x_i)  \notin \j_i)},
    \label{eq:N2}
\end{equation}
where $n$ is the number of samples in the dataset and $NN(\x_i)$ is the nearest neighbour of a given sample $\x_i$. Note that $NN(\x_i)$ can either belong to the same class of $\x_i$ ($NN(\x_i)  \in \j_i$) or not ($NN(\x_i)  \notin \j_i$). 

This measure of feature separability, however, has limitations related to the shape of the samples distribution; also, in our case, $d(\x_i, NN(\x_i)  \notin \j_i)$ is frequently equal to zero, meaning that for several samples the nearest \emph{enemy} (sample from a different class) is overlapped with the sample itself. Therefore, we decided to focus on the denominator of \eqref{eq:N2}, which is also known as the Local Set Radius (LSR), i.e., the radius of the hypershpere centered in one sample and tangent to the nearest enemy:% (see Figure \ref{fig:LSR-example}).

\begin{equation}
    LSR = d(\x_i, NN(\x_i)  \notin \j_i)
\end{equation}

In Figure \ref{fig:LSR}a it is possible to observe that the average LSR is closer to zero for chains having Twitter as their last sharing step (right-most part of the graph), thus suggesting a lower separability of such classes.  %Low average LSRs are also registered for chains having Twitter in $\C[-1]$, i.e., the penultimate step. 

To further highlight this, Figure~\ref{fig:LSR}b--d reports the LSR values aggregated in subset related to the specific platforms, for the {\DCT} features ({\META} exhibits the same trend and {\HEADER} is rather uninformative as the LSR is equal to zero in the majority of cases). Figure~\ref{fig:LSR}d demonstrates how the presence of Twitter in the sharing chain affects the average LSR: in fact, all groups have values lower than $\Omega^*$, which contains all classes that do not include Twitter in $\C[0]$ or $\C[-1]$. Moreover, in Figure~\ref{fig:LSR}b--c, it is possible to see how the lowest values of LSR for chains having Facebook or Flicker in $\C[-1]$ do occur when Twitter is in $\C[0]$.

From this analysis it is clear that, while Twitter is perfectly recognizable when occurring as the last sharing step, chains that contain Twitter in $\C[0]$ or $\C[-1]$ are not separable with the employed sets of features. In general, we can state that Twitter is particularly disruptive with regard to the forensic traces left by the previous sharing platforms, at least for the set of traces considered in this work.  

The detection of Twitter at a generic step $F_{-\ell}$ of the reconstruction pipeline should therefore be regarded as a stopping point, given the unacceptable performance of previous sharing detectors. Accordingly, we designed an \emph{informed} version of the cascade architecture that stops the reconstruction process when it encounters Twitter, as discussed in the following final section.

% *** I-CASCADE ***

\subsection{Informed framework}

The informed framework only differs from the standard system described in Section~\ref{sec:multi} by the introduction of an additional stopping condition. 

As formulated in \eqref{eq:block}, a backtracking block $F_{-\ell}$ interrupts the reconstruction process when it receives from the previous step a chain of length $\ell-1$, meaning that the end of the chain has been reached. A second stopping condition is introduced by the BKS fusion modules, which may reject input samples that fall outside of the learned distribution. 

In the informed framework, we simply modify \eqref{eq:block} by introducing an additional condition:
\begin{equation}
    \Fell(\x,\C) = 
    \begin{cases}
        \C & \text{if } \C \in \Omega_{\ell-1}\\
        & \text{or } \C[-(\ell-1)]=\lab{TW}{}{}\\
        f^\C_{-\ell}(\x) & \text{otherwise}
    \end{cases}
\end{equation}
This way, when $F_{-\ell}$ receives a chain that contains Twitter in $\C[-(\ell-1)]$, i.e., the last detected step, the reconstruction stops. Clearly, this modification results in a reduced number of classifiable sharing chains. 

In the specific implementation evaluated in this work (with $L=3$ backtracking blocks), all chains of the form $\lab{*}{*}{TW}$ and $\lab{*}{TW}{*}$ collapse in the classes $\lab{TW}{}{}$ and $\lab{TW}{*}{}$, respectively, thus obtaining 21 classes at the output of $F_{-2}$.

\begin{figure}[t]
    \centering
    \includegraphics[width=\columnwidth]{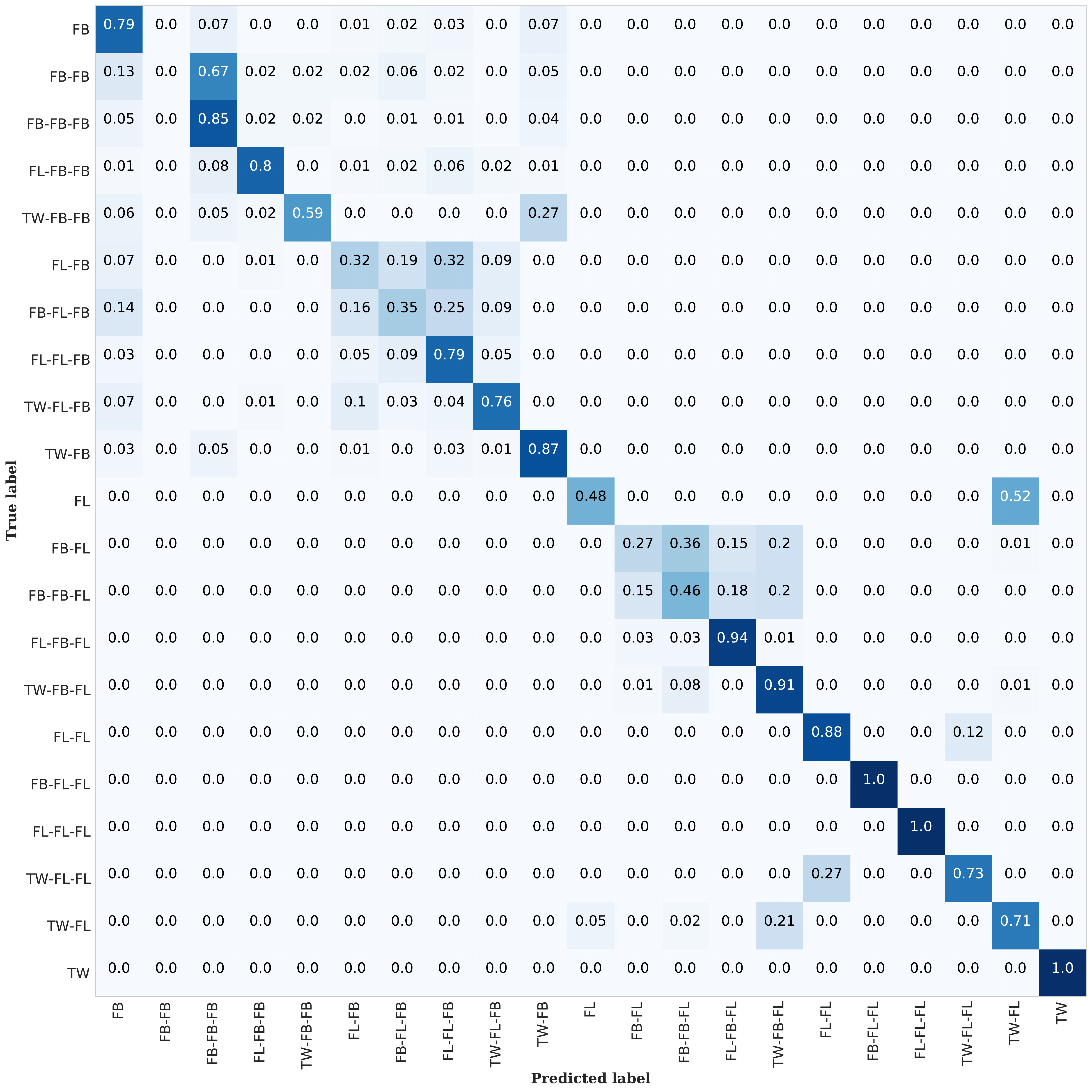}
    \caption{Detection performance of backtracking block $F_{-2}$ in the informed cascade system, which identifies up to the last three platforms in the sharing chain, ($\C[-2],\C[-1],\C[0]$); test carried out with the fusion of all three feature descriptors (see Table~\ref{tab:i-cascade} for the results with different feature subsets); note that all chains of the form $(\textrm{*},\textrm{*},\textrm{TW})$ and $(\textrm{*},\textrm{TW},\textrm{*})$ are condensed in 
    TW and $(\textrm{TW},\textrm{*})$, respectively.}
    \label{fig:i-step3}
\end{figure}

Table~\ref{tab:i-cascade} reports the overall accuracy values and rejection rates for each step of the informed system (note that $F_0$ is not affected by the modification), while
Figure~\ref{fig:i-step3} shows the final 21-by-21 confusion matrix in output of the $F_{-2}$ backtracking block of the informed system.

\setlength{\tabcolsep}{4.5pt}
\begin{table}[t]
    \centering
        \caption{Per-step performance of the informed cascade system.}
    \begin{tabular}{rcccccc}
         \toprule    
         & \multicolumn{2}{c}{$F_0$} & \multicolumn{2}{c}{$F_{-1}$} & \multicolumn{2}{c}{$F_{-2}$}\\
         & \multicolumn{2}{c}{3 classes} & \multicolumn{2}{c}{9 classes} & \multicolumn{2}{c}{21 classes}\\
         \midrule
        {\scriptsize Single classifiers} & ACC &  & ACC &  & ACC & \\
         \midrule
         {\scriptsize {\DCT}} & 0.8634 &  & 0.6219 & & 0.5104 & \\
         {\scriptsize {\META}} & 0.9296 &  & 0.7302 & & 0.6141 &\\
         {\scriptsize {\HEADER}} & \textbf{1.0000} &  & 0.8046 & & 0.5623 &\\
         {\scriptsize Random guess} & 0.3333 & & 0.1111 & & 0.0476 & \\
         \midrule
         {\scriptsize Fused classifiers} & ACC & REJ & ACC & REJ & ACC & REJ\\
         \midrule
         {\tiny {\DCT}+{\META}} & 0.9296 & 0.0000 & 0.7496 & 0.0000 & 0.6399 & 0.0056\\
         {\tiny {\META}+{\HEADER}} & \textbf{1.0000} & 0.0000 & 0.9011 & 0.0000 & 0.7774 & 0.0000\\
         {\tiny {\DCT}+{\META}+{\HEADER}} & \textbf{1.0000} & 0.0000 & \textbf{0.9057} & 0.0010 & \textbf{0.8105} & 0.0377\\
         \bottomrule
    \end{tabular}
    \label{tab:i-cascade}
\end{table}

In Figure~\ref{fig:i-step3} we can observe how confusions typically occur when the same platform is concatenated multiple times: $\lab{FB}{FB}{}$ gets mistaken for $\lab{FB}{FB}{FB}$; $\lab{FL}{FB}{}$ and $\lab{FB}{FL}{}$ are confused with $\lab{FL}{FL}{FB}$ and $\lab{FB}{FB}{FL}$, respectively; the only notable exception is $\lab{FL}{}{}$ being confused with $\lab{TW}{FL}{}$.

With the informed cascade, accuracy values are significantly higher than with the standard framework, reaching 90\% at step $F_{-1}$ and 81\% at step $F_{-2}$, with the fusion of all three descriptors. More importantly, rejection rates are dramatically reduced, especially at step $F_{-2}$, confirming that most rejections were due to the non-separability of Twitter-related sub-chains. 

\section{Conclusions}\label{sec:conc}
The possibility to reverse engineer the history of a digital content in terms of sharing operations can represent a valuable resource in tracing perpetrators of deceptive visual contents, thus playing a significant role in preserving the trustworthiness of digital media and countering misinformation effects.
In this work we addressed the media recycling scenario, with the purpose of reconstructing the sharing history of images through multiple uploads on different social media platforms. We proposed a framework that allows the fusion of heterogeneous feature descriptors, combining them in a cascaded classification system that identifies one step of the sharing chain at a time. Also, we introduced a novel set of container-based features extracted from the header of the image file. 
Experiments demonstrated that the combination of content- and container-based features outperforms the single classifiers in the identification of the various sharing steps. As a side result, we observed by means of a feature separability analysis that uploads on Twitter turn out to be particularly disruptive towards the traces left by previous platforms, at least for the considered forensic features, thus hampering the reconstruction of the sharing chain. Taking this into account, and therefore interrupting the process at the first detection of an upload on Twitter, the reconstruction achieves an overall 81\% accuracy for chains of up to three steps. 
Given the flexibility of the proposed method, future works may extend the presented architecture with additional platforms, allowing to reconstruct more complex and diversified sharing chains.

% *** APPENDICES ***

% \appendices

% \section{}

\section*{Acknowledgment}
We thank Prof. Fabio Roli (University of Cagliari, Italy) for the valuable insights on classifier fusion and the BKS method. We also thank Chiara Albisani (University of Florence, Italy) for contributing to the parsing of header data. 

This material is based upon work supported by the Defense Advanced Research Projects Agency (DARPA) under Agreement No. HR00112090136 and by the PREMIER project, funded by the Italian Ministry of Education, University, and Research (MIUR).

\ifCLASSOPTIONcaptionsoff
  \newpage
\fi



%\bibliographystyle{IEEEtran}
%\begin{thebibliography}{1}

%\end{thebibliography}

\bibliographystyle{IEEEtran}
\bibliography{references}

\begin{thebibliography}{10}
\providecommand{\url}[1]{#1}
\csname url@samestyle\endcsname
\providecommand{\newblock}{\relax}
\providecommand{\bibinfo}[2]{#2}
\providecommand{\BIBentrySTDinterwordspacing}{\spaceskip=0pt\relax}
\providecommand{\BIBentryALTinterwordstretchfactor}{4}
\providecommand{\BIBentryALTinterwordspacing}{\spaceskip=\fontdimen2\font plus
\BIBentryALTinterwordstretchfactor\fontdimen3\font minus
  \fontdimen4\font\relax}
\providecommand{\BIBforeignlanguage}[2]{{%
\expandafter\ifx\csname l@#1\endcsname\relax
\typeout{** WARNING: IEEEtran.bst: No hyphenation pattern has been}%
\typeout{** loaded for the language `#1'. Using the pattern for}%
\typeout{** the default language instead.}%
\else
\language=\csname l@#1\endcsname
\fi
#2}}
\providecommand{\BIBdecl}{\relax}
\BIBdecl

\bibitem{smith2019}
K.~Smith, ``126 amazing social media statistics and facts,''
  \url{https://www.brandwatch.com/blog/amazing-social-media-statistics-and-facts/},
  2019.

\bibitem{statista2021}
{Statista Research Department}, ``Hours of video uploaded to youtube every
  minute as of may 2019,''
  \url{https://www.statista.com/statistics/259477/hours-of-video-uploaded-to-youtube-every-minute/},
  2021.

\bibitem{thies2016face}
J.~Thies, M.~Zollhofer, M.~Stamminger, C.~Theobalt, and M.~Niessner,
  ``Face2face: Real-time face capture and reenactment of rgb videos,'' in
  \emph{Proceedings of the IEEE Conference on Computer Vision and Pattern
  Recognition (CVPR)}, June 2016.

\bibitem{ngo2021self}
M.~Ngo, S.~Karaoglu, and T.~Gevers, ``Self-supervised face image manipulation
  by conditioning gan on face decomposition,'' \emph{IEEE Transactions on
  Multimedia}, pp. 1--1, 2021.

\bibitem{allen2017}
T.~C. G.~Allen, ``Artificial intelligence and national security,'' \emph{Belfer
  Center Study}, 2017.

\bibitem{Verdoliva2020media}
L.~Verdoliva, ``Media forensics and deepfakes: an overview,'' \emph{IEEE
  Journal of Selected Topics in Signal Processing, in press}, 2020.

\bibitem{pasquini2021media}
C.~Pasquini, I.~Amerini, and G.~Boato, ``Media forensics on social media
  platforms: a survey,'' \emph{EURASIP Journal on Information Security}, vol.
  2021, no.~1, pp. 1--19, 2021.

\bibitem{cheung2015connection}
M.~Cheung, J.~She, and Z.~Jie, ``Connection discovery using big data of
  user-shared images in social media,'' \emph{IEEE Transactions on Multimedia},
  vol.~17, no.~9, pp. 1417--1428, 2015.

\bibitem{mada2019trust}
B.~E. Mada, M.~Bagaa, and T.~Taleb, ``Trust-based video management framework
  for social multimedia networks,'' \emph{IEEE Transactions on Multimedia},
  vol.~21, no.~3, pp. 603--616, 2019.

\bibitem{moltisanti2015image}
M.~Moltisanti, A.~Paratore, S.~Battiato, and L.~Saravo, ``Image manipulation on
  facebook for forensics evidence,'' in \emph{International Conference on Image
  Analysis and Processing}, 2015, pp. 506--517.

\bibitem{caldelli2017}
R.~{Caldelli}, R.~{Becarelli}, and I.~{Amerini}, ``Image origin classification
  based on social network provenance,'' \emph{IEEE Transactions on Information
  Forensics and Security}, vol.~12, no.~6, pp. 1299--1308, 2017.

\bibitem{amerini2017}
I.~{Amerini}, T.~{Uricchio}, and R.~{Caldelli}, ``Tracing images back to their
  social network of origin: A {CNN}-based approach,'' in \emph{IEEE Workshop on
  Information Forensics and Security (WIFS)}, 2017, pp. 1--6.

\bibitem{caldelli2018}
R.~{Caldelli}, I.~{Amerini}, and C.~T. {Li}, ``{PRNU}-based image
  classification of origin social network with {CNN},'' in \emph{European
  Signal Processing Conference (EUSIPCO)}, 2018, pp. 1357--1361.

\bibitem{amerini2019social}
I.~Amerini, C.-T. Li, and R.~Caldelli, ``Social network identification through
  image classification with {CNN},'' \emph{IEEE Access}, vol.~7, pp.
  35\,264--35\,273, 2019.

\bibitem{mazumdar2019}
A.~Mazumdar, J.~Singh, Y.~S. Tomar, and P.~K. Bora, ``Detection of image
  manipulations using siamese convolutional neural networks,'' in \emph{Pattern
  Recognition and Machine Intelligence}, 2019, pp. 226--233.

\bibitem{mullan2019}
P.~Mullan, C.~Riess, and F.~Freiling, ``Forensic source identification using
  jpeg image headers: The case of smartphones,'' \emph{Digital Investigation},
  vol.~28, pp. S68 -- S76, 2019.

\bibitem{giudice2017}
O.~Giudice, A.~Paratore, M.~Moltisanti, and S.~Battiato, ``A classification
  engine for image ballistics of social data,'' in \emph{Image Analysis and
  Processing - ICIAP 2017}, 2017.

\bibitem{phan2018}
Q.~{Phan}, C.~{Pasquini}, G.~{Boato}, and F.~G.~B. {De Natale}, ``Identifying
  image provenance: An analysis of mobile instant messaging apps,'' in
  \emph{IEEE International Workshop on Multimedia Signal Processing (MMSP)},
  2018, pp. 1--6.

\bibitem{phan2019}
Q.~{Phan}, G.~{Boato}, R.~{Caldelli}, and I.~{Amerini}, ``Tracking multiple
  image sharing on social networks,'' in \emph{IEEE International Conference on
  Acoustics, Speech and Signal Processing (ICASSP)}, 2019, pp. 8266--8270.

\bibitem{siddiqui2019}
N.~{Siddiqui}, A.~{Anjum}, M.~{Saleem}, and S.~{Islam}, ``Social media origin
  based image tracing using deep {CNN},'' in \emph{2019 Fifth International
  Conference on Image Information Processing (ICIIP)}, 2019, pp. 97--101.

\bibitem{ih2016}
C.~Pasquini, P.~Sch\"{o}ttle, R.~B\"{o}hme, G.~Boato, and
  F.~F.~P\`{e}rez-Gonz\`{a}lez, ``Forensics of high quality and nearly
  identical {JPEG} image recompression,'' in \emph{ACM Information Hiding and
  Multimedia Security Workshop}, Vigo, Galicia, Spain, 2016, pp. 11--21.

\bibitem{pevny2008detection}
T.~Pevny and J.~Fridrich, ``Detection of double-compression in jpeg images for
  applications in steganography,'' \emph{IEEE Transactions on Information
  Forensics and Security}, vol.~3, no.~2, pp. 247--258, 2008.

\bibitem{farid2009}
E.~Kee, M.~K. Johnson, and H.~Farid, ``Digital image authentication from jpeg
  headers,'' \emph{IEEE Transactions on Information Forensics and Security},
  vol.~6, no.~3, pp. 1066--1075, 2011.

\bibitem{yang2020efficient}
P.~Yang, D.~Baracchi, M.~Iuliani, D.~Shullani, R.~Ni, Y.~Zhao, and A.~Piva,
  ``Efficient video integrity analysis through container characterization,''
  \emph{IEEE Journal of Selected Topics in Signal Processing}, vol.~14, no.~5,
  pp. 947--954, 2020.

\bibitem{exiftool}
P.~Harvey, ``Exiftool,'' \url{https://exiftool.org/}, 2021.

\bibitem{Ruta2000}
D.~Ruta and B.~Gabrys, ``An overview of classifier fusion methods,''
  \emph{Computing and Information systems}, vol.~7, no.~1, pp. 1--10, 2000.

\bibitem{Moreno2006}
F.~Moreno-Seco, J.~M. Inesta, P.~J.~P. De~Le{\'o}n, and L.~Mic{\'o},
  ``Comparison of classifier fusion methods for classification in pattern
  recognition tasks,'' in \emph{Joint IAPR International Workshops on
  Statistical Techniques in Pattern Recognition (SPR) and Structural and
  Syntactic Pattern Recognition (SSPR)}.\hskip 1em plus 0.5em minus 0.4em\relax
  Springer, 2006, pp. 705--713.

\bibitem{Huang1993}
Y.~S. Huang and C.~Y. Suen, ``The behavior-knowledge space method for
  combination of multiple classifiers,'' in \emph{IEEE computer society
  conference on computer vision and pattern recognition}.\hskip 1em plus 0.5em
  minus 0.4em\relax Institute of Electrical Engineers Inc (IEEE), 1993, pp.
  347--347.

\bibitem{Huang1995}
------, ``A method of combining multiple experts for the recognition of
  unconstrained handwritten numerals,'' \emph{IEEE transactions on pattern
  analysis and machine intelligence}, vol.~17, no.~1, pp. 90--94, 1995.

\bibitem{Raudys2003}
{\v{S}}.~Raudys and F.~Roli, ``The behavior knowledge space fusion method:
  Analysis of generalization error and strategies for performance
  improvement,'' in \emph{International Workshop on Multiple Classifier
  Systems}.\hskip 1em plus 0.5em minus 0.4em\relax Springer, 2003, pp. 55--64.

\bibitem{dang2015raise}
D.-T. Dang-Nguyen, C.~Pasquini, V.~Conotter, and G.~Boato, ``Raise: A raw
  images dataset for digital image forensics,'' in \emph{Proceedings of the 6th
  ACM multimedia systems conference}, 2015, pp. 219--224.

\bibitem{ho2002data}
T.~K. Ho, ``A data complexity analysis of comparative advantages of decision
  forest constructors,'' \emph{Pattern Analysis \& Applications}, vol.~5,
  no.~2, pp. 102--112, 2002.

\end{thebibliography}


%\end{IEEEbiography}


%\newpage



%\vfill


\end{document}